\documentclass[a4paper,11pt]{article}
\usepackage{graphicx}
\usepackage{epstopdf}
\usepackage{amsmath,amssymb,amsfonts}
\usepackage{epsfig,cancel,cite}

\newlength{\xtrawidth}
\newlength{\xtraheight}
\newcommand{\be}{\begin{equation}}
\newcommand{\ee}{\end{equation}}
\newcommand{\beq}{\begin{equation}}
\newcommand{\eeq}{\end{equation}}
\newcommand{\ba}{\begin{array}}
\newcommand{\ea}{\end{array}}
\newcommand{\bea}{\begin{eqnarray}}
\newcommand{\eea}{\end{eqnarray}}
\newcommand{\bean}{\begin{eqnarray*}}
\newcommand{\eean}{\end{eqnarray*}}
\newcommand{\eref}[1]{(\ref{#1})}

\newcommand{\comment}[1]{}

\def\fnote#1#2{\begingroup\def\thefootnote{#1}\footnote{#2}
     \addtocounter{footnote}{-1}\endgroup}

\begin{document}

\vspace{1cm}

\title{{\Large \bf Two Hundred Heterotic Standard Models \\ on Smooth Calabi-Yau Threefolds}}

\vspace{2cm}

\author{
Lara B. Anderson${}^{1}$,
James Gray${}^{2}$,
Andre Lukas${}^{3}$,
Eran Palti${}^{4}$
}
\date{}
\maketitle
\begin{center} {\small ${}^1${\it Department of Physics, University of
      Pennsylvania, \\ Philadelphia, PA 19104-6395, U.S.A.}\\[0.2cm]
      ${}^2${\it Arnold-Sommerfeld-Center for Theoretical Physics, \\
       Department f\"ur Physik, Ludwig-Maximilians-Universit\"at M\"unchen,\\
       Theresienstra\ss e 37, 80533 M\"unchen, Germany}\\[0.2cm]
       ${}^3${\it Rudolf Peierls Centre for Theoretical Physics, Oxford
       University,\\
       $~~~~~$ 1 Keble Road, Oxford, OX1 3NP, U.K.\\[0.2cm]
       ${}^4${\it Centre de Physique Theorique, Ecole Polytechnique, CNRS, 91128 Palaiseau, France,}}}\\

\fnote{}{andlara@physics.upenn.edu,~james.gray@uni-muenchen.de,~lukas@physics.ox.ac.uk,~eran.palti@cpht.polytechnique.fr} 

\end{center}

\abstract{\noindent }We construct heterotic standard models by
compactifying on smooth Calabi-Yau three-folds in the presence of
purely Abelian internal gauge fields. A systematic search over
complete intersection Calabi-Yau manifolds with less than six K\"ahler
parameters leads to over 200 such models which we present. Each of
these models has precisely the matter spectrum of the MSSM, at least
one pair of Higgs doublets, the standard model gauge group and no
exotics.  For about 100 of these models there are four
additional U(1) symmetries which are Green-Schwarz anomalous
and, hence, massive. In the remaining cases, three U(1) symmetries are
anomalous while the fourth, massless one can be spontaneously broken
by singlet vacuum expectation values. The presence of additional
global U(1) symmetries, together with the possibility of switching on
singlet vacuum expectation values, leads to a rich phenomenology which
is illustrated for a particular example. Our database of standard
models, which can be further enlarged by simply extending the
computer-based search, allows for a detailed and systematic
phenomenological analysis of string standard models, covering issues
such as the structure of Yukawa couplings, R-parity violation, proton
stability and neutrino masses.  \newpage
\tableofcontents

%
%

\section{Introduction}

There is a long history of attempting to construct four dimensional
theories, from smooth compactifications of the heterotic string, with
a matter sector which precisely matches that of the Minimal
Supersymmetric Standard Model (MSSM). Indeed, the subject of string
phenomenology started in this way in the 1980s when various attempts
were made to build models based upon the ``standard embedding''. In
that approach, the gauge bundle was taken to be the holomorphic
tangent bundle, with $SU(3)$ structure group, or deformations of the
tangent bundle \cite{Green:1987mn}. In recent years, more general
gauge configurations have been used in an attempt to achieve
phenomenologically viable physics. Slope-stable\footnote{Slope-stable
  bundles satisfy the Hermitian-Yang-Mills equations required for
  $N=1$ supersymmetry in $4$-dimensions \cite{Green:1987mn}.} bundles
with $SU(n)$ structure groups (for $n=3,4,5$), unrelated to the
tangent bundle, have been used in the attempt to build stringy
standard
models~\cite{Bouchard:2005ag,Braun:2005ux,Braun:2005bw,Braun:2005zv,Braun:2005nv,Bouchard:2006dn,Blumenhagen:2006ux,Blumenhagen:2006wj,Anderson:2007nc,Anderson:2008uw,Anderson:2009mh}. These constructions were based upon the use of non-Abelian gauge
field configurations on smooth Calabi-Yau three-folds.\footnote{Another class of models are based on non-smooth CY orbifolds, these have been shown to also allow for an appropriate massless spectrum as well as other phenomemological features \cite{Buchmuller:2005jr,Buchmuller:2006ik,Lebedev:2006kn,Kim:2007mt,Lebedev:2008un,Nibbelink:2009sp,Blaszczyk:2009in,Blaszczyk:2010db,Kappl:2010yu}. There are also constructions based on non-geometric settings such as the free-fermionic models as studied in \cite{Assel:2009xa,Christodoulides:2011zs,Cleaver:2011ir}.}

In this paper we adopt a different approach to constructing standard
models in smooth Calabi-Yau three-fold compactifications of heterotic
string and M-theory. Instead of using the non-Abelian constructions
mentioned in the preceding paragraph, we shall construct models where
the gauge field configuration in the internal dimensions is simply a
sum of line bundles - that is a set of $U(1)$ fluxes. This is the
extremal form of the so-called ``split" or reducible bundles first
studied in Refs.~\cite{He:2003tj,Blumenhagen:2005ga}.

There are two key aspects to this approach that differentiate it from
the traditional non-Abelian one. The first is a practical one: it is
much simpler to construct, and calculate the resulting spectrum of,
Abelian bundles than non-Abelian ones. As a result, an algorithmic and
systematic approach to such (heterotic) string model building is
relatively straightforward and can be used to analyse vast numbers of
line bundle sums over Calabi-Yau manifolds. Rather than attempting to
fine-tune the construction of a single example, this large data set
can be scanned for realistic models, using methods of computational
algebraic geometry\footnote{Similar scans for non-Abelian
  constructions have been started in
  Refs.~\cite{Anderson:2007nc,Anderson:2008uw,Anderson:2009mh,He:2009wi}
  and further results will be presented in Ref.~\cite{monad_scan}.}
\cite{cicypackage,Gray:2008zs,singular}. This paper presents our
first results from an investigation along these lines. We have
systematically scanned line bundle sums on Calabi-Yau three-folds
(defined as complete intersections in products of projective spaces)
with Hodge number $h^{1,1}(X)\leq 5$ and have found 208 heterotic
standard models. It is important to note that these models are all
``global" in that they correspond to explicit Calabi-Yau threefolds
and holomorphic vector bundles leading to fully consistent heterotic
theories. All 208 models have the precise matter spectrum of the MSSM,
at least one pair of Higgs doublets, the standard model gauge group
and no exotics charged under the standard model group of any kind. The
number of models constructed should be considered with the knowledge
that to date, only $3$ other smooth heterotic standard models have
been produced in the
literature~\cite{Bouchard:2005ag,Braun:2005ux,Anderson:2009mh}.

The second key aspect of heterotic line bundle model building is
related to additional $U(1)$ symmetries. We will consider line bundle
sums with structure group $S(U(1)^5)$ whose commutant within $E_8$ is
$SU(5)\times S(U(1)^5)\cong SU(5)\times U(1)^4$. Hence, before Wilson
line breaking, our models are based on $SU(5)$ GUT theories with four
additional $U(1)$ symmetries. Phenomenologically, the vector bosons
associated with those $U(1)$ symmetries should of course be
massive. Fortunately, there are two mechanisms to generate such
masses, both within our control. The first is the Green-Schwarz
mechanism: the $U(1)$ vector bosons can acquire a large mass, close to
the compactification scale, due to a gauging of axion shift
symmetries. For 105 of our 208 models this happens for all four $U(1)$
symmetries, so that the low-energy gauge group is precisely that of
the standard model. The remaining models have three anomalous and,
hence, massive $U(1)$ symmetries while the fourth Abelian gauge factor
remains massless, as long as the internal bundle is a sum of line
bundles. In this case, we can invoke the second mechanism, namely
moving away from the split locus in bundle moduli space such that the
bundle structure group becomes non-Abelian, thus removing the extra
$U(1)$ from the low energy gauge group. In the effective field-theory
this amounts to giving supersymmetric vacuum expectation values (VEVs)
to bundle moduli fields. We have explicit control over the spectrum of
such bundle moduli and can, therefore, analyse this effect in detail.

Another important physical implication, which is tied to the above
discussion, is that the Green-Schwarz anomalous $U(1)$ symmetries give
rise to residual $U(1)$ global symmetries in the effective
theory. These global symmetries impose constraints on the possible
operators present in the theory and may forbid problematic operators
such as those that lead to proton decay or R-parity violation. They
may also serve as Froggatt-Nielsen type symmetries to explain the
patterns of observed quark and lepton masses. This interplay between
$U(1)$ symmetries, their spontaneous breaking through bundle moduli
VEVs, and the resulting operators in the low energy theory, leads to a
rich arena for phenomenology \cite{Kuriyama:2008pv,Anderson:2010tc}.

In this paper, we present the physical ideas behind our work, the
database of 208 standard models, and an exploration some of the
phenomenological issues by focusing on a particular example. A more
comprehensive study will be presented in a forthcoming
paper~\cite{bigpaper}.

The plan of this paper is as follows. In the next section we briefly explain the basic model-building set-up. Section~\ref{gs} reviews the Green-Schwarz mechanism and its particular implications for our models. In section \ref{lb} we describe our scanning procedure and its main results. As an illustrative example, one of our standard models is presented in Section~\ref{eg}. Section \ref{eran} discusses the phenomenological implications of the anomalous $U(1)$ symmetries and bundle moduli VEVs in more detail, focusing on the particular example introduced earlier. We present a brief summary and an outlook in Section \ref{conc}. The data for all 208 standard models is listed in the Appendix. 

\section{Model building set-up}
We consider compactifications of the $E_8\times E_8$ heterotic string on a smooth Calabi-Yau three-fold, $X$, with a freely acting discrete symmetry, $\Gamma$. In practice, we will use complete intersection Calabi-Yau manifolds (CICYs) which are defined as the common zero locus of homogeneous polynomials in an ambient product of projective spaces $\mathbb{P}^{n_1}\times \dots\times\mathbb{P}^{n_m}$. These manifolds have been classified~\cite{Gagnon:1994ek,the_cy_list} and their freely-acting symmetries are known~\cite{Braun:2010vc}. In the present paper, we will explore all CICYs with freely acting symmetries and Hodge number satisfying $h^{1,1}(X)\leq 5$. It turns out, all these manifolds are ``favourable" in the sense that $h^{1,1}(X)=m$, so that their whole second cohomology is spanned by the restrictions of the K\"ahler forms, $J_i$, of the ambient projective spaces. Line bundles, $L$, on $X$, the main building blocks of our bundle construction, can hence be denoted as $L={\cal O}_X({\bf k})$, where ${\bf k}$ is an $m$--dimensional integer vector such that $c_1({\cal O}_X({\bf k}))=k^iJ_i$. 

As mentioned earlier, on $X$ we consider vector bundles $V$ with structure group $S(U(1)^5)$, that is, sums of line bundles
\begin{equation}
  V=\bigoplus_{a=1}^5 L_a\; \mbox{ where }\; L_a={\cal O}_X({\bf k}_a)\; , \label{Vdef}
 \end{equation}
 satisfying
 \begin{equation}
  c_1(V)=\sum_{a=1}^5c_1^i(L_a)J_i=0\; .\label{c10}
 \end{equation} 
Hence, for a given three-fold, $X$, and a given symmetry, $\Gamma$, a model is specified by the $5\,h^{1,1}(X)$ integers $k_a^i$. In our model scan, we will restrict ourselves to bundles, $V$, for which
\begin{equation}\label{anom_canc}
c_2(TX)-c_2(V)=[C]~~~,~~[C]~\text{an effective class in}~H_2(X,\mathbb{Z})
\end{equation}
which allows for an anomaly-free supersymmetric completion by addition of an appropriate number of five-branes wrapping $C$. Supersymmetry conditions on the bundle $V$ itself will be discussed in the next section.
 
The structure group is embedded into $E_8$ via the sub-group chain $S\left(U(1)^5\right)\subset SU(5)\subset E_8$, so that the four-dimensional gauge group, before Wilson line breaking, is the GUT group $SU(5)\times S(U(1)^5)$. In general, the low-energy theory contains the standard $SU(5)$ multiplets ${\bf 10}$, $\bar{\bf 5}$ (and their conjugates) and bundle moduli singlets ${\bf 1}$.  In addition, the above multiplets are labeled by $S(U(1)^5)$ charges, which can be represented as integers vectors ${\bf q}=(q_1,\ldots ,q_5)$. Due to the unit determinant condition in $S(U(1)^5)$, two such charge vectors ${\bf q}$ and $\tilde{\bf q}$ have to be identified if ${\bf q}-\tilde{\bf q}\in\mathbb{Z}(1,1,1,1,1)$ and, as a result, each charge vector with five same entries corresponds to the trivial representation. This fact will be of importance later on when we discuss the $S(U(1)^5)$ invariant operators in the four-dimensional effective theory. With this notation, the matter multiplet content of the GUT group is
\begin{equation}
 {\bf 10}_{{\bf e}_a},\;\bar{\bf 10}_{-{\bf e}_a},\;\bar{\bf 5}_{{\bf e}_a+{\bf e}_b},\; {\bf 5}_{-{\bf e}_a-{\bf e}_b},\;{\bf 1}_{{\bf e}_a-{\bf e}_b},\;{\bf 1}_{-{\bf e}_a+{\bf e}_b}\; ,
 \end{equation}
where $a<b$. Here, the subscripts are $S(U(1)^5)$ charges with ${\bf e}_a$ the $a^{\rm th}$ standard unit vector in five dimensions. These multiplets are associated to particular line bundle cohomology groups, as summarised in Table~\ref{tab1}, and their numbers can be determined by computing the dimensions of these cohomology groups. 
\begin{table}[h]
\small
\begin{center}
\begin{tabular}{|l|c|c|c|}\hline
 multiplet&$S(U(1)^5)$ charge&associated line bundle $L$&contained in\\\hline\hline
 ${\bf 10}_{{\bf e}_a}$&${\bf e}_a$&$L_a$&$V$\\\hline
 $\bar{\bf 10}_{-{\bf e}_a}$&$-{\bf e}_a$&$L_a^*$&$V^*$\\\hline
 $\bar{\bf 5}_{{\bf e}_a+{\bf e}_b}$&${\bf e}_a+{\bf e}_b$&$L_a\otimes L_b$&$\wedge^2V$\\\hline
 ${\bf 5}_{-{\bf e}_a-{\bf e}_b}$&$-{\bf e}_a-{\bf e}_b$&$L_a^*\otimes L_b^*$&$\wedge^2V^*$\\\hline
 ${\bf 1}_{{\bf e}_a-{\bf e}_b}$&${\bf e}_a-{\bf e}_b$&$L_a\otimes L_b^*$&$V\otimes V^*$\\
 ${\bf 1}_{-{\bf e}_a+{\bf e}_b}$&$-{\bf e}_a+{\bf e}_b$&$L_a^*\otimes L_b$&\\\hline
 \end{tabular}
\parbox{6in}{\caption{\it\small Multiplet content, charges and associated line bundles of the $SU(5)\times S(U(1)^5)$ GUT theory. The indices $a,b,\ldots$ are in the range $1,\dots ,5$ and ${\bf e}_a$ denotes the standard five-dimensional unit vector in the $a^{\rm th}$ direction. The number of each type of multiplet is obtained from the first cohomology, $H^1(X,L)$,  of the associated line bundle $L$.}\label{tab1}}
\end{center}
\end{table} 
For CICYs, line bundle cohomology can be explicitly computed by applying the methods described in Refs~\cite{Anderson:2008uw,Anderson:2008ex,Anderson:2009mh}. Compared to a standard $SU(5)$ GUT theory, the multiplet content of our models is split into sub-sectors, labeled by different $S(U(1)^5)$ charges. Invariance under $S(U(1)^5)$ restricts the allowed operators in the low-energy theory and this will be of importance for the phenomenological discussion later on. In particular, we note that the bundle moduli singlets carry non-trivial $S(U(1)^5)$ charges, so operators involving these singlets are constrained as well. This leads to an interesting interplay between $S(U(1)^5)$ invariance and switching on singlet VEVs. In the language of vector bundles, non-zero singlet VEVs corresponds to moving away from the Abelian locus in bundle moduli space to a bundle with non-Abelian structure group. 

The further breaking of the GUT theory to the standard model proceeds in the standard way via Wilson lines. For the bundle $V$ to descend to the quotient Calabi-Yau manifold, $X/\Gamma$, it has to be equivariant under the symmetry $\Gamma$ \cite{Donagi:2003tb}, a property which can be explicitly checked for line bundles using the methods described in Ref.~\cite{Anderson:2009mh}. Note that for an equivariant line bundle, $L$, the cohomology groups $H^i(X,L)$ form representations under the group $\Gamma$. A Wilson line on the quotient, pointing into the standard hypercharge direction then breaks the GUT group into the standard model group times the massive $S(U(1)^5)$ symmetry. Let us consider a standard model multiplet with Wilson line representation $R_W$ which originates from a GUT multiplet with associated line bundle, $L$. The number of these multiplets can be computed from the $\Gamma$ invariant part of $H^1(X,L)\otimes R_W$. In essence, once the GUT multiplet content is known, computing the particle content after Wilson line breaking is a matter of applying representation theory of the finite group $\Gamma$. 

\section{Additional $U(1)$ symmetries and Green-Schwarz mechanism}\label{gs}
We turn now to the fate of the four additional $U(1)$ symmetries in $S(U(1)^5)\cong U(1)^4$ which arise in our models. The Green-Schwarz mechanism in heterotic theories has been understood for many years (see \cite{Lukas:1999nh} and \cite{Blumenhagen:2005ga,Anderson:2009sw,Anderson:2009nt,Anderson:2010ty} for some recent papers on the subject). It is known that Abelian factors in the bundle structure group give rise to a gauging of certain axion shift symmetries in the four dimensional effective theory. In our context, for each line bundle, $L^a$, in $V$, the K\"ahler axions, $\chi^i$, the supersymmetric partners of the K\"ahler moduli, $t^i$, acquire the following transformation\footnote{The equations below receive a one loop correction due to a non-trivial shift of the dilatonic and $M5$-brane axions. This has been explicitly studied in Ref.~\cite{Blumenhagen:2005ga,Anderson:2009nt} but will be neglected in the present context as it does not affect our discussion.}
\begin{equation} \label{mrgauging}
 \delta \chi^i=-c_1^i(L_a)\eta_a\; ,
\end{equation}
with transformation parameter $\eta_a$. Note that, from Eq.~\eqref{c10}, only four of these transformation, corresponding to the four $U(1)$ symmetries, are independent. Each such transformation leads to a D-term which schematically reads
\begin{equation}
 D_a=\frac{\mu(L_a)}{\kappa}-\sum_IQ_{aI}|C_I|^2\; . \label{Dterms}
\end{equation}
Here, $\kappa=d_{ijk}t^it^jt^k$ is the K\"ahler moduli space
pre-potential with the triple intersection numbers $d_{ijk}$ of $X$
and $C_I$ are matter fields and bundle moduli with charges $Q_{aI}$
under $S(U(1)^5)$. The slope, $\mu(L_a)$, of the line bundle $L_a$ is
defined as
\begin{equation}
 \mu(L_a)=c_1^i(L_a)\kappa_i\; \mbox{ with } \; \kappa_i=d_{ijk}t^jt^k\; .
\end{equation} 
We can now discuss the conditions on the line bundle sum $V$ arising
from $N=1$ supersymmetry. From a four-dimensional point of view, for a
supersymmetric vacuum, all D-terms~\eqref{Dterms} must vanish. The
locus in bundle moduli space where $V$ is split into a sum of line
bundles corresponds to setting all VEVs of the fields $C_I$ to
zero. Hence, all slopes, $\mu(L_a)$, must vanish simultaneously,
somewhere in K\"ahler moduli space. This is, of course, the well-known
condition for line bundle sums to preserve supersymmetry.  For the
equations $k^i_a\kappa_i=0$ to have a non-trivial solution it must be
the case that the
\begin{equation}
 (\mbox{number of lin. independent }{\bf k}_a) < h^{1,1}(X)\; . \label{c1cons}
\end{equation}
This implies strong model building constraints for Calabi-Yau manifolds with a small Hodge number $h^{1,1}(X)$ and explains why we were not able to find standard models on CICYs with $h^{1,1}(X)=2,3$.

At the split locus in bundle moduli space, the mass matrix for the $S(U(1)^5)$ vector bosons is given by
\begin{equation}
 M_{ab}=G_{ij}c_1^i(L_a)c_1^j(L_b)\; ,
\end{equation}
where $G_{ij}=-\partial_i\partial_j\ln\kappa$ is the K\"ahler moduli space metric of $X$. Since $G_{ij}$ is positive definite, the number of massless $U(1)$ vector fields must equal $4-{\rm rank}(k_a^i)$ and can, hence, be easily determined from the integers $k_a^i$ which specify our models. Combining this statement with the inequality~\eqref{c1cons} we learn that
\begin{equation}
 (\mbox{number of massless }U(1)\mbox{ vector fields})>4-h^{1,1}(X)\; .
\end{equation} 
Hence, for Calabi-Yau manifolds with $h^{1,1}(X)=4$ at least one massless $U(1)$ vector field remains, while $h^{1,1}(X)=5$ is the smallest Hodge number for which all $U(1)$ vector fields can receive masses from the Green-Schwarz mechanism.

\section{Searching for line bundle standard models}\label{lb}

Our scanning procedure involves the following basic steps. For a given Calabi-Yau manifold $X$ with freely-acting Abelian symmetry, $\Gamma$, we generate a large number of line bundle sums, $V=\bigoplus_{a=1}^5L_a$, satisfying $c_1(V)=0$, each specified by an integer matrix $k_a^i=c_1^i(L_a)$. In practice, we restrict the entries $k_a^i$ to run in a certain finite range. In a first filtering step, we extract all line bundle sums which are supersymmetric (that is, all slope conditions $\mu(L_a)=0$ can be satisfied for some K\"ahler parameters of $X$) and which satisfy \eref{anom_canc}. This ensures that all remaining models give rise to consistent heterotic vacua on X. Subsequently, we extract all line bundle sums which are equivariant under $\Gamma$, so that the model can be quotiented by $\Gamma$. 

The second step involves imposing physical constraints on the spectrum
of the $SU(5)\times S(U(1)^5)$ GUT theory. These conditions can be
easily inferred from Table~\ref{tab1}. First we impose that
$h^1(X,V)=3|\Gamma|$ and $h^1(X,V^*)=0$, where $|\Gamma|$ is the order
of the discrete symmetry group $\Gamma$. This is to ensure that
downstairs we have precisely three $SU(5)$ families of ${\bf 10}$
multiplets and no $\bar{\bf 10}$ anti-families. As can easily be
proved, it then follows that
$h^1(X,\wedge^2V)-h^1(X,\wedge^2V^*)=3|\Gamma|$ so that there is a
downstairs chiral asymmetry of three $\bar{\bf 5}$ families. Secondly,
we need at least one vector-like $\bar{\bf 5}$--${\bf 5}$ pair in
order to retain a pair of Higgs doublets so we also require that
$h^1(X,\wedge^2V^*)>0$.

With these conditions imposed we have a model with the standard model
gauge group (times four $U(1)$ symmetries, some or all massive), three
families of quarks and leptons and whatever remains from the $\bar{\bf
  5}$--${\bf 5}$ pair. To increase the chance that the Higgs triplets
can be removed we demand that $h^1(L_a^*\otimes L_b^*)<|\Gamma|$ for
all $a<b$, so that the number of such pairs is smaller than the group
order in each sector. In this case, it can be shown that for
appropriate choices of equivariant structure and Wilson line, for all
$208$ models, the Higgs triplets can be projected out and at least one
pair of Higgs doublets can be kept \cite{bigpaper}.

As a first step, the above procedure has been carried out for all CICYs with symmetries and $h^{1,1}(X)\leq 5$ in the standard list~\cite{the_cy_list}. We recall that $h^{1,1}(X)=5$ is the smallest value for which all four additional $U(1)$ symmetries can become massive due to the Green-Schwarz mechanism, so it is sensible to scan up to this Hodge number at least. For the 6 CICYs with $h^{1,1}(X)=2$ this has been done for line bundle entries in the range $-10\leq k_a^i\leq 10$ and for the 12 CICYs with $h^{1,1}(X)=3$ the range $-3\leq k_a^i\leq 3$ has been covered. No model passing all the above tests has been found. As indicated earlier, this can be traced back to the stability constraint~\eqref{c1cons} which is particularly strong for low $h^{1,1}(X)$. 

The $19$ CICYs with symmetries at $h^{1,1}(X)=4$ have been scanned in the range $-3\leq k_a^i\leq 3$ and $28$ models passing all tests have been found. The scan over the $23$ CICYs with $h^{1,1}(X)=5$ in the range $-2\leq k_a^i\leq 2$ resulted in $180$ models. Altogether, we have found $208$ heterotic line bundle standard models which are explicitly listed in the Appendix. For $105$ of these models, all for $h^{1,1}(X)=5$, all additional $U(1)$ symmetries are Green-Schwarz anomalous and super-heavy. For the remaining models we have three anomalous, massive $U(1)$ symmetries and one massless one. As indicated earlier, this remaining $U(1)$ can be easily broken spontaneously by switching on singlet VEVs and, for this reason, these models have been included.

These results have been obtained from a scan over roughly $10^{12}$ integer matrices $k_a^i$ generated initially. Hence, a ``one in a billion" rule of thumb \cite{Gmeiner:2005vz} is not too far from the truth in this part of the heterotic vacuum space. It should be mentioned that this task has not required high performance computing but was completed (within several weeks) on a standard desktop machine. Extending to larger ranges for the $k_a^i$ and to CICYs with larger $h^{1,1}(X)$ is merely a question of computing power. 

\section{A standard model example} \label{eg}

In order to illustrate our result and to set up a more explicit
context for the subsequent phenomenological discussion, we will now
present one of our $208$ standard models in more detail. This will be
sufficient for the main purpose of this paper which is to merely
indicate the rich structure of model building possibilities. A
detailed analysis for all standard models in our database will be
carried out in a forthcoming paper~\cite{bigpaper}.

Our example lives on the $h^{1,1}(X)=5$ CICY with configuration matrix
\bea \label{cy} X= \left( \ba{c|cccc} \mathbb{P}^1 & 1 & 1 & 0 & 0 \\
  \mathbb{P}^1 & 0 & 0 & 0 & 2 \\ \mathbb{P}^1 & 0 & 0 & 2 & 0 \\
  \mathbb{P}^1 & 2 & 0 & 0 & 0 \\ \mathbb{P}^3 & 1 & 1 & 1 & 1 \ea
\right)^{5,37}_{-64} \eea
defined in the ambient space $(\mathbb{P}^1)^{\times 4}\times \mathbb{P}^3$, as indicated in the first column of the configuration matrix. We denote the homogeneous coordinates of the four $\mathbb{P}^1$ by $x_{i\alpha}$, where $i=1,2,3,4$ and $\alpha=0,1$ and the $\mathbb{P}^3$ coordinates by $y_\alpha$, where $\alpha=0,\ldots ,3$. The remaining columns of the above matrix specify the multi-degrees of four homogeneous polynomials on the ambient space whose common zero locus defines the CICY, $X$. The subscript is the Euler number and the superscripts provide the Hodge numbers $h^{1,1}(X)$ and $h^{2,1}(X)$, which count the number of K\"ahler and complex structure moduli, respectively. The second cohomology of $X$ is spanned by the five ambient space K\"ahler forms $J_i$ and the cone of allowed K\"ahler forms $J=t^iJ_i$ is specified by $t^i>0$ for all $i$. The triple intersection numbers of $X$ have the following non-zero components (as well as those related by symmetry of the indices)
\bea d_{123}=d_{124}= d_{134}= d_{234}=
d_{235}&=&2\nonumber \\ d_{125}= d_{135}= d_{145}=
d_{245}= d_{255}= d_{345} = d_{355}&=&4 \\ \nonumber
d_{155}= d_{455} = d_{555}&=&8\; . \eea
The second Chern class of the tangent bundle is $c_2(TX)=(24,24,24,24,56)$, relative to a basis of four-forms dual to the ambient space K\"ahler forms $J_i$. The manifold is simply connected but can be divided by a freely acting $\Gamma=\mathbb{Z}_2$ symmetry which transforms the ambient space coordinates as
\bea \label{sym}
(x_{i0},x_{i1}) \to (-x_{i0},x_{i1}) \; , \quad (y_0,y_1,y_2,y_3) \to (-y_0,-y_1,y_2,y_3)
\eea
Our model is specified by the sum of line bundles
\bea V\; =\; \bigoplus_{a=1}^5L_a &=&{\cal O}_X(1,0,0,-1,0) \oplus {\cal  O}_X(1,-1,-2,0,1) \oplus {\cal O}_X(0,1,1,1,-1)\oplus\nonumber\\
&&{\cal O}_X(0,-1,1,0,0)_X \oplus {\cal O}_X(-2,1,0,0,0)\; .
 \label{bundle}
\eea
This bundle satisfies $c_1(V)=0$ and \eref{anom_canc}. In addition, using the above intersection numbers, it can be verified that the slope conditions $\mu(L_a)=0$ can be simultaneously satisfied at a locus in the K\"ahler cone of $X$. It can also be verified that $V$ is $\mathbb{Z}_2$ equivariant and, hence, descends to a bundle on the ``downstairs" quotient space $X/\mathbb{Z}_2$. The bundle~\eqref{bundle} has four linearly independent Chern classes $c_1(L_a)={\bf k}_a$. From our earlier discussion this means that all four additional $U(1)$ symmetries are Green-Schwarz anomalous and, hence, massive. Consequently, the downstairs gauge group is precisely the standard model gauge group.

The non-vanishing cohomology groups of the constituent line bundles $L_a$ are given by
\begin{equation}
 h^1(X,L_2)=4\; ,\quad h^1(X,L_5)=2\; .
\end{equation} 
We recall from Table~\ref{tab1} that the cohomology groups $H^1(X,L_a)$ count the number of GUT multiplets ${\bf 10}_{{\bf e}_a}$. Hence, after dividing by the symmetry order, $|\Gamma|=2$, this leads to three multiplets, ${\bf 10}_{{\bf e}_2}$,  ${\bf 10}_{{\bf e}_2}$, ${\bf 10}_{{\bf e}_5}$, in the downstairs spectrum. 

The non-vanishing first cohomology groups of tensor products $L_a\otimes L_b$ and $L_a^*\otimes L_b^*$ are
\begin{equation} 
 h^1(X,L_2\otimes L_4)=4\; ,\quad h^1(X,L_4\otimes L_5)=2\; ,\quad h^1(X,L_2\otimes L_5)=1\; ,\quad h^1(X,L_2^*\otimes L_5^*)=1\; .
\end{equation}
From Table~\ref{tab1}, the cohomology groups $H^1(X,L_a\otimes L_b)$ and $H^1(X,L_a^*\otimes L_b^*)$ count the number of $\bar{\bf 5}_{{\bf e}_a+{\bf e}_b}$ and ${\bf 5}_{-{\bf e}_a-{\bf e}_b}$ GUT multiplets, respectively. This means downstairs we have three multiplets, $\bar{\bf 5}_{{\bf e}_2+{\bf e}_4}$, $\bar{\bf 5}_{{\bf e}_2+{\bf e}_4}$, $\bar{\bf 5}_{{\bf e}_4+{\bf e}_5}$ plus whatever remains from the vector-like pair of $\bar{\bf 5}_{{\bf e}_2+{\bf e}_5}$ and ${\bf 5}_{-{\bf e}_2-{\bf e}_5}$ multiplets after Wilson line breaking. It turns out, in line with general arguments above, that both Higgs triplets can be projected out while the pair of Higgs doublets can be kept. As a result, the complete spectrum of multiplets charged under the standard model group is precisely that of the MSSM, as summarised in Table~\ref{tab2} below.
\begin{table}[!h]
\small
\begin{center}
\begin{tabular}{|l||l|l|l|l|l|l|l|l|}\hline
name&${\bf 10}_1$&${\bf 10}_2$&${\bf 10}_3$&$\bar{\bf 5}_1$&$\bar{\bf 5}_2$&$\bar{\bf 5}_3$&$H_u$&$H_d$\\\hline
$S(U(1)^5)$ charge&${\bf e}_2$&${\bf e}_2$&${\bf e}_5$&${\bf e}_2+{\bf e}_4$&${\bf e}_2+{\bf e}_4$&${\bf e}_4+{\bf e}_5$&$-{\bf e}_2-{\bf e}_5$&${\bf e}_2+{\bf e}_5$\\\hline
\end{tabular}
\parbox{6in}{\caption{\it\small Charges of the standard model multiplets in our example model. Each multiplet arises with multiplicity one. For simplicity, families are denoted by $SU(5)$ representations but should be thought of as broken up into standard model multiplets, keeping the $S(U(1)^5)$ charge unchanged.}\label{tab2}}
\end{center}
\end{table}
From Table~\ref{tab1}, the number of singlets ${\bf 1}_{{\bf e}_a-{\bf e}_b}$ is determined by $H^1(X,L_a\otimes L_b^*)$. For our model, the non-vanishing first cohomology groups in this sector are
\begin{equation}
\begin{array}{llll} 
 h^1(X,L_2\otimes L_1^*)=4\; ,& h^1(X,L_5\otimes L_1^*)=8\; ,& h^1(X,L_2\otimes L_3^*)=4\; ,& h^1(X,L_2\otimes L_4^*)=12\\
 h^1(X,L_2\otimes L_5^*)=11\; ,&h^1(X,L_5\otimes L_2^*)=3\; ,&h^1(X,L_4\otimes L_5^*)=6\; .&
\end{array}
\end{equation} 
After Wilson line breaking, this gives rise to seven types of singlets, denoted by $C_1,\ldots , C_7$, whose charges and multiplicities are listed in Table~\ref{tab3}.
\begin{table}[!h]
\small
\begin{center}
\begin{tabular}{|l||l|l|l|l|l|l|l|}\hline
name&$C_1$&$C_2$&$C_3$&$C_4$&$C_5$&$C_6$&$C_7$\\\hline
$S(U(1)^5)$ charge&${\bf e}_2-{\bf e}_1$&${\bf e}_5-{\bf e}_1$&${\bf e}_2-{\bf e}_3$&${\bf e}_2-{\bf e}_4$&${\bf e}_2-{\bf e}_5$&
${\bf e}_5-{\bf e}_2$&${\bf e}_4-{\bf e}_5$\\\hline
multiplicity&2&4&2&6&5&1&3\\\hline
\end{tabular}
\parbox{6in}{\caption{\it\small Charges and multiplicities for the seven types of bundle moduli singlets in our example model.}\label{tab3}}
\end{center}
\end{table}

To summarise, our example model has the exact spectrum and gauge group of the MSSM, plus seven types of bundle moduli fields which are singlets under the standard model group. All those fields carry charges under the remnant global $S(U(1)^5)$ symmetry which constrains the four-dimensional effective theory. The phenomenology resulting from the interplay between this global symmetry and switching on VEVs for the singlet fields will be discussed in the next section. 

\section{Residual symmetries and singlet VEVs} \label{eran}

In the previous section we presented an example from our standard model database which has exactly the matter spectrum of the MSSM along with some gauge singlet fields. In this model, all four additional $U(1)$ symmetries are Green-Schwarz anomalous, so that their associated gauge bosons are super-heavy and, hence, absent from the low-energy theory. However, they leave behind global $U(1)$ symmetries (see \cite{Kuriyama:2008pv,Anderson:2010tc} for recent explorations of such symmetries in heterotic theories) which allow us to constrain the operator spectrum of the theory~\cite{Kuriyama:2008pv} and push the phenomenological study beyond the mere computation of the spectrum. Similar considerations apply to the other standard models in our database. In this section, we would like to discuss some of these phenomenological issues in general and illustrate our points within the context of the example model. A systematic study for all models will be presented in a forthcoming paper~\cite{bigpaper}. We also note that the themes presented in this section are recurrent within the F-theory GUT literature, see for example \cite{Heckman:2009mn,Marsano:2009wr,Dudas:2009hu,Dudas:2010zb,Marsano:2010sq}.

The study of allowed operators in the theory involves finding
$S(U(1)^5)$ invariant field combinations. We recall that $S(U(1)^5)$
charges are labeled by integer vectors ${\bf q}=(q_1,\ldots,q_5)$ and,
as a result of the determinant one condition in $S(U(1)^5)$, two such
integer vectors ${\bf q}$ and $\tilde{\bf q}$ have to be identified if
${\bf q}-\tilde{\bf q}\in\mathbb{Z}(1,1,1,1,1)$. A particular operator
is therefore allowed if its charge vector is entirely zero or if it is
non-zero but with all entries equal. For our example, the explicit
charge vectors of the MSSM fields and the seven singlet fields $C_I$
are given in Tables~\ref{tab2} and \ref{tab3}. We note that these
charges are not flavour-universal, a feature which is generic for
heterotic line bundle models\footnote{Note that this shows that the
  approach adopted in \cite{Dudas:2009hu} within the F-theory
  framework of allowing different families to come from different
  matter curves is in fact rather generic.}. In our analysis, we also
allow the singlets $C_I$ to develop a VEV~\footnote{So long as this
  VEV remains small compared to the compactification scale, we can
  define a valid perturbative theory near the Abelian locus in moduli
  space. For more details on the mass scales associated to these VEVs,
  see \cite{Anderson:2009sw}.} which we denote by
\begin{equation}
 \epsilon_I=\langle C_I\rangle\; .
\end{equation} 
As a result, the allowed terms involve higher dimension operators with
singlet insertions - much like in the Froggatt-Nielsen setup
\cite{Froggatt:1978nt}. As mentioned earlier, $S(U(1)^5)$ gauge bosons
which did not receive a mass from the Green-Schwarz mechanism can
become massive due to the spontaneous breaking induced by these
VEVs. This is the reason why we have included such models in our list
of $208$ standard models given in the Appendix. In the following, we
will frequently write down operators in terms of $SU(5)$ GUT
multiplets, for simplicity. This is appropriate because every standard
model field within a given $SU(5)$ multiplet carries the same
$S(U(1)^5)$ charge. However, we should keep in mind that, even though
we use the language of $SU(5)$ GUTs, the subsequent discussion applies
to heterotic standard models.

It is important to note that an operator allowed by the $S(U(1)^5)$
symmetries is not necessarily present in the theory - this would
require further calculations to determine \cite{Anderson:2009ge}. In
particular, the theory might have further discrete symmetries which
forbid some operators allowed by $S(U(1)^5)$. However, an $S(U(1)^5)$
non-invariant operator is definitely forbidden at the perturbative
level. It can still be generated by non-perturbative effects but one
would expect such a contribution to be suppressed.

\subsection{Proton decay}

One of the strongest constraints on supersymmetric theories comes from dimension four proton decay, induced by  superpotential operators of the form ${\bf \bar{5}}\, {\bf \bar{5}}\, {\bf 10}$ with matter multiplets $\bar{\bf 5}$ and ${\bf 10}$. In our context, these operators can be written as ${\bf \bar{5}}_{{\bf e}_a+{\bf e}_b} {\bf \bar{5}}_{{\bf e}_c+{\bf e}_d} {\bf 10}_{{\bf e}_f}$ and, hence, have a total $S(U(1)^5)$ charge ${\bf e}_a+{\bf e}_b+{\bf e}_c+{\bf e}_d+{\bf e}_f$. Such an operator is allowed precisely if all five charge vectors involved are different in which case the total charge is $(1,1,1,1,1)$. Whether this happens depends on the precise charges of the matter fields and has to be analysed in detail for each of our standard models. For our example, the matter field charges in Table~\ref{tab2} show that all such operators are forbidden and, hence, this particular model is safe from dimension four proton decay at the Abelian split locus. What happens if we move away from this locus by switching on singlet VEVs $\epsilon_I$? In this case, we have to worry about re-creating such operators by singlet VEV insertions. Again this is a matter of detailed analysis for each given model, but for our example model the singlet charges in Table~\ref{tab3} show that they are never re-created for any number of singlet insertions. Our example model is therefore safe from dimension four proton decay in at least a neighbourhood of the Abelian locus in bundle moduli space. 

A less-constrained but nevertheless important effect is dimension five proton decay, induced by operator of the form
${\bf \bar{5}}_{{\bf e}_a+{\bf e}_b} {\bf 10}_{{\bf e}_c}{\bf 10}_{{\bf e}_d} {\bf 10}_{{\bf e}_f}$ with total charge ${\bf e}_a+{\bf e}_b+{\bf e}_c+{\bf e}_d+{\bf e}_f$. For our example, such operators are forbidden, as the $S(U(1)^5)$ charges in Table~\ref{tab2} show and, from the singlet charges in Table~\ref{tab3}, they are not re-created by singlet insertions.

The above results regarding proton decay are promising. However within
our models, forbidding proton decay using the $S(U(1)^5)$ symmetry
comes at a price. From the neutrality of the Yukawa couplings in the
MSSM, it is easy to show that the only $U(1)$ symmetry that can forbid
dimension five proton decay is one that is not vector-like for the up-
and down-Higgs. Such a symmetry is often referred to as a Pecci-Quinn
symmetry, $U(1)_{PQ}$. In our example, the Higgs pair is indeed
vector-like and so there is no $U(1)_{PQ}$. The reason for the absence
of dimension five proton decay in this model is that, as discussed
below, the down-type Yukawa couplings are forbidden by $S(U(1)^5)$
and, hence, the standard MSSM reasoning based on the presence of these
couplings does not apply.  Of course, this may not be a real problem
as the down-type Yukawa couplings may be generated by non-perturbative
effects. Such non-perturbative effects may or may not re-introduce
proton decay. Whether or not this occurs can be decided at the present
level of sophistication, relying on the information provided by the
$S(U(1)^5)$ symmetry, by writing down the relevant gauge invariant
non-perturbative contributions to the theory given the axion
transformations \eqref{mrgauging}. As with the perturbative terms being
discussed in this section, whether or not such terms actually appear
in the theory, as opposed to simply being allowed by gauge invariance,
requires more detailed calculation to determine.

In fact, we find that, under fairly general assumptions, the issue
discussed in the proceeding paragraph is generic in heterotic line
bundle standard models. Assuming that the low-energy spectrum does not
contain exotic states, such as Higgs triplets, Higgs pairs are always
vector-like under $S(U(1)^5)$ and, hence, there is no $U(1)_{PQ}$
symmetry. The underlying model-building reasons for this will be
discussed in Ref.~\cite{bigpaper}. Here, we present a more intuitive
argument which follows from anomaly cancellation. The key observation
is that, since the Green-Schwarz couplings only depend on the gauge
field-strength, the GUT-breaking Wilson-line cannot affect
Green-Schwarz anomaly cancellation. Considering the mixed anomalies of
two standard model gauge factors with one of the additional $U(1)$
symmetries, together with the MSSM matter spectrum, these can only
match the GUT anomalies if the Higgs fields are vector like under the
$U(1)$ symmetry. Consequently, there is either no $U(1)_{PQ}$ symmetry
or the theory contains exotic matter fields~\footnote{There is a very
  similar story in F-theory, for which we note our Wilson-line
  argument above also applies, in the case of hypercharge flux
  doublet-triplet splitting
  \cite{Marsano:2009wr,Dudas:2010zb,Marsano:2010sq,Dolan:2011iu}. Also
  note that this argument applies to an unbroken $U(1)_{PQ}$ and can
  be evaded by having an approximate symmetry, i.e. breaking it well
  below the cutoff scale.}.

\subsection{R-parity violation}

There is a set of superpotential operators which violate the MSSM
R-parity and which lead to too large neutrino masses, namely operators
of the form ${\bf 5}^{H_u}_{-{\bf e}_a-{\bf e}_b} {\bf \bar{5}}_{{\bf
    e}_c+{\bf e}_d}$ with $S(U(1)^5)$ charge $-{\bf e}_a-{\bf
  e}_b+{\bf e}_c+{\bf e}_d$. For our example, an inspection of the
charges in Table~\ref{tab2} shows that these operators are
forbidden. This is consistent with our cohomology calculation which
shows that, at the Abelian split locus, the three $\bar{\bf 5}$ matter
multiplets and the up-Higgs are indeed massless. However, the
dimension four operator $C_3\bar{\bf 5}_3H_u$ is allowed so it is
possible to induce some of these R-parity violating terms by switching
on a VEV for $C_3$. To be safe we have to demand that
$\epsilon_3=\langle C_3\rangle =0$ and this is sufficient to remove
all similar operators with any number of singlet insertions.

\subsection{$\mu$--term}
\label{muterm}
A related discussion applies to the $\mu$-term, $\mu H_uH_d$. As we have argued above, for our models Higgs doublets come in vector-like pairs under the $S(U(1)^5)$ symmetry. Consequently, the $\mu$-term is allowed by $S(U(1)^5)$. However, as the cohomology calculation shows, all our 208 standard models have at least one massless Higgs pair at the Abelian locus in bundle moduli space. Hence, for all these models, the $\mu$-term is absent from the superpotential for reasons unrelated to the $S(U(1)^5)$ symmetry. What happens when we move away from the Abelian locus by switching on singlet VEVs? A quick glance at Table~\ref{tab3} shows that our example model has no singlets which are completely uncharged under $S(U(1)^5)$, so dimension four terms of the form $C_IH_uH_d$ are forbidden. In fact, this is generic for all our models. Bundle moduli with charge ${\bf e}_a-{\bf e}_b$ are counted by the first cohomology of $L_a\otimes L_b^*$. Singlets under $S(U(1)^5)$ can only arise for $a=b$ but in this case $H^1(X,L_a\otimes L_a^*)=H^1(X,{\cal O}_X)=0$.

As a result, the lowest dimension at which a $\mu$-term can be generated is five. The relevant operators are of the form $C_IC_JH_uH_d$ where $C_I$ and $C_J$ need to have opposite $S(U(1)^5)$ charge. For sufficiently small VEVs, $\epsilon_I$, $\epsilon_J$, this can provide a string-theoretical realisation of the solution to the $\mu$-problem proposed in Ref.~\cite{Kim:1994eu}. In our example model such a dimension five operator, $C_5C_6H_uH_d$, is allowed and, if indeed present, could give rise to a $\mu$-term of an acceptable size provided the product $\epsilon_5\epsilon_6$ is sufficiently small. A small value for this product is independently suggested by the pattern of up-type Yukawa couplings discussed below. 

\subsection{Yukawa couplings}

Three possible types of contributions to the (superpotential) Yukawa coupling arise in our models. First we have regular dimension four  Yukawa couplings. In the up sector they are of the form ${\bf 5}^{H_u}_{-{\bf e}_a-{\bf e}_b}{\bf 10}_{{\bf e}_c}{\bf 10}_{{\bf e}_d}$ and allowed provided ${\bf e}_a+{\bf e}_b={\bf e}_c+{\bf e}_d$. The down sector Yukawa couplings, $\bar{\bf 5}^{H_d}_{{\bf e}_a+{\bf e}_b}\bar{\bf 5}_{{\bf e}_c+{\bf e}_d}{\bf 10}_{{\bf e}_f}$, are allowed if ${\bf e}_a+{\bf e}_b+{\bf e}_c+{\bf e}_d+{\bf e}_f=(1,1,1,1,1)$. As we have mentioned earlier, the $S(U(1)^5)$ symmetry is not flavour-universal, so this generates a pattern of order one entries in the Yukawa matrices. Further contributions, proportional to the VEVs $\epsilon_I$ or products thereof, can be generated by vacuum insertions once singlet VEVs are switched on. This amounts to a string-theoretical realisation of a Froggatt-Nielsen~\cite{Froggatt:1978nt} type model for fermion masses.\footnote{We note that, as show in \cite{Dudas:2009hu}, the group theory of $E_8$ allows for an accurate recreation of the observed masses and mixing of the quarks and leptons.} Finally, we may have non-perturbative contributions. Here, we will only consider the first two types of effects explicitly and we stress that they can be straightforwardly analysed for all our standard models.

However, when discussing the results, we should keep in mind that non-perturbative corrections to Yukawa couplings are rather common in string theory and provide a possible mechanism to generate small fermion masses. It is, therefore, not absolutely necessary to explain the full structure of Yukawa couplings from a Froggatt-Nielsen approach based on the $S(U(1)^5)$ symmetry. However, we should certainly require that the top Yukawa coupling is generated perturbatively at order one.  

For our example model, the charges in Table~\eqref{tab2} show that, in the absence of singlet VEVs, the up-type Yukawa matrix has rank two while the down-type Yukawa matrix vanishes identically. Switching on VEVs $\epsilon_5=\langle C_5\rangle$ and $\epsilon_6=\langle C_6\rangle$ the Yukawa matrices take the form
\be
Y^U = \left( \begin{array}{ccc} \epsilon_{5} & 1 & 1 \\ 1 & \epsilon_{6} & \epsilon_{6} \\ 1 & \epsilon_{6} & \epsilon_{6} \end{array} \right) \;,\quad\quad
Y^D = \left( \begin{array}{ccc} 0 & 0 & 0 \\ 0 & 0 & 0 \\ 0 & 0 & 0\end{array} \right) \;.
\ee
Note that order one coefficients have been omitted so that $Y^U$ generically has rank three. The eigenvalues of $Y^U$ are of order $1$, $1$ and $\epsilon_6$, giving two heavy generations and one potentially lighter one, depending on the position in moduli space. The down-type Yukawa couplings are vanishing and so require non-perturbative effects in order to be generated.

Generally, when giving VEVs to singlets we must ensure that supersymmetry is preserved, that is, the D-terms~\eqref{Dterms} must remain zero. This is a  very mild restriction as the K\"ahler moduli can adjust themselves to minimise the D-term potential for many choices of singlet VEVs. For our example, it is even simpler to prove the existence of VEVs compatible with supersymmetry. Since $C_5$ and $C_6$ are vector-like we can set $\epsilon_5=\epsilon_6$ and keep the K\"ahler moduli fixed, so that the FI and matter field contributions to the D-term vanish independently.\footnote{In order to recreate the hierarchy between the top and up quark masses, and solve the D-terms, we should take $\epsilon_5=\epsilon_6\sim10^{-6}$ which interestingly implies the $\mu$-term operator discussed in section \ref{muterm} is naturally at the TeV scale.} 

It is worth noting a practical advantage originating from the $S(U(1)^5)$ symmetry, in relation to the physical Yukawa couplings in heterotic compactifications. It is generally very difficult to calculate the structure of the kinetic terms of the matter fields and so deducing the physical Yukawa couplings from the holomorphic ones is non-trivial. The additional $U(1)$ symmetries can be of help in this regard because they can restrict the matter field kinetic terms severely.

\subsection{Neutrino physics}

The bundle moduli serve as good candidates for right-handed neutrinos \cite{Kuriyama:2008pv}. For our example model, we can consider the fields $C_4$ as forming the right handed neutrinos. In this case we have the superpotential operators, in GUT field notation,
\be
W \supset {\bf 5}_{H_u} {\bf \bar{5}}_3 C_4 + \epsilon_6 {\bf 5}_{H_u} {\bf \bar{5}}_2 C_4 + \epsilon_6 {\bf 5}_{H_u} {\bf \bar{5}}_1 C_4 + \left(\epsilon_6 \epsilon_7\right)^2 C_4 C_4 \;.
\ee
The first three terms provide Dirac neutrinos masses while the last gives a Majorana mass to $C_4$ thereby realising the see-saw mechanism. However note that there is also a possible linear term $\epsilon_6 \epsilon_7 C_4$ which must be forbidden in some way (in the MSSM this is done using matter-parity).

\section{Conclusions and outlook} \label{conc}

In this paper, we have presented a database of $208$ heterotic standard models based on smooth Calabi-Yau manifolds and Abelian bundles over them. All of these models have the precise matter spectrum of the MSSM, one or more pairs of Higgs doublets, the standard model gauge group with possibly one additional $U(1)$ symmetry and no exotic matter charged under the standard model of any kind. For $105$ of these models, there is no additional $U(1)$ symmetry so that the gauge group is exactly the standard model group. For the remaining models this $U(1)$ can be spontaneously broken by switching on singlet VEVs. We have presented an example model from our database with the exact gauge group and spectrum of the MSSM in more detail. 

An interesting additional feature of our heterotic line bundle models is the presence of a global, flavour non-universal $S(U(1)^5)\cong U(1)^4$ symmetry which restricts the structure of the four-dimensional effective theory. Standard model fields as well as bundle moduli singlets are charged under $S(U(1)^5)$. The interplay between this symmetry and switching on singlet VEVs, thereby moving away from a purely Abelian bundle, provides a rich phenomenological setting for issues such as proton stability, R-parity violation, the $\mu$-problem and fermion masses. We have discussed some of these issues and have illustrated them with our example. It turns out, in this model, that the $S(U(1)^5)$ symmetry stabilises the proton, allows for an order one top Yukawa coupling, facilitates a possible solution to the $\mu$-problem and may provide a realisation of the see-saw mechanism for neutrino masses. 

We believe that our results raise the phenomenology of heterotic
Calabi-Yau compactifications to a new level. Phenomenological problems
beyond the calculation of the spectrum can now be addressed within a
sizable class of quasi-realistic explicit models, rather than for a small
number of individual models which are likely to fail more
sophisticated phenomenological requirements. Such a systematic
phenomenological analysis, for the standard models presented here,
will be carried out in a forthcoming paper~\cite{bigpaper}.

Our work can be extended in a variety of ways. Scans over CICYs with Hodge numbers $h^{1,1}>5$ and larger ranges of bundles are underway and are likely to lead to more standard models. It would be interesting to perform a similar scan for heterotic line bundle models on Calabi-Yau hypersurfaces in toric varieties, as classified in Ref.~\cite{Kreuzer:2000qv, Kreuzer:2000xy}, although this requires developing a number of technical tools \cite{Kreuzer:2002uu, Blumenhagen:2010pv}.

\section*{Acknowledgments}
E.~P.~would like to thank Angel Uranga for correspondence and the
University of Oxford for hospitality during the completion of the
work. A.~L.~is supported by the EC 6th Framework Programme
MRTN-CT-2004-503369 and would like to thank Ecole Polytechnique for
hospitality while part of this work was being carried out. The work of
EP was supported by the European ERC Advanced Grant 226371 MassTeV and
the PITN contract PITN-GA-2009-237920. L.~A. is supported in part by
the DOE under Grant No. DE-AC02-76-ER-03071. J.~G. would like to
acknowledge support by the NSF-Microsoft grant NSF/CCF-1048082.

\appendix
\section{Line bundle standard models on $h^{1,1}(X)=4,5$ CICYs}
In this Appendix we provide tables with all $208$ line bundle standard models  which we have found on CICYs with $h^{1,1}(X)=4,5$. The scan has been performed over all line bundle sums $V=\bigoplus_{i=1}^5{\cal O}_X({\bf k}_a)$ with entries in the range $-3\leq k_a^i\leq 3$ for $h^{1,1}(X)=4$ and $-2\leq k_a^i\leq 2$ for $h^{1,1}(X)=5$. The methodology and the general results of this scan have already been described in Section~\ref{lb}.

The notation in the tables is as follows. The first row contains information about the CICY, namely the CICY identifier (that is, its position in the standard CICY list~\cite{the_cy_list}), the standard configuration matrix with the Euler number as sub-script and $h^{1,1}(X),h^{2,1}(X)$ as super-scripts and the freely acting symmetry by which the model is divided. Each subsequent table entry specifies a line bundle sum by providing the five vectors ${\bf k}_a$. As explained in Section~\ref{gs}, the number of massless $U(1)$ symmetries at the Abelian locus in bundle moduli space is given by $4$ minus the number of linearly independent vectors ${\bf k}_a$ and can, hence, be directly read of from the data provided here. 
\begin{table}[!h]
\begin{center}
\scriptsize
\begin{tabular}{|l|l|}
\hline
CICY 6784: {\tiny $\left(\begin{array}{l}\mathbb{P}^1\\\mathbb{P}^1\\\mathbb{P}^1\\\mathbb{P}^3\end{array}\right|\left.\begin{array}{lll}1&1&0\\0&0&2\\2&0&0\\1&1&2\\\end{array}\right)^{4,36}_{-64}$}&$\mathbb{Z}_2\times\mathbb{Z}_2$\\\hline\hline
(3,2,-2,-1)(1,-1,0,0)(-1,0,1,0)(-1,0,1,0)(-2,-1,0,1)&(2,2,1,-1)(1,-1,0,0)(1,-1,0,0)(-1,0,-2,1)(-3,0,1,0)\\\hline
(2,1,0,-1)(0,1,-3,0)(0,-2,1,1)(-1,0,1,0)(-1,0,1,0)&(2,1,-3,0)(0,1,2,-1)(0,-2,-1,1)(-1,0,1,0)(-1,0,1,0)\\\hline
(1,0,-1,0)(1,0,-1,0)(1,-2,0,1)(0,1,2,-1)(-3,1,0,0)&(1,2,2,-1)(1,0,-3,0)(0,-1,1,0)(0,-1,1,0)(-2,0,-1,1)\\\hline
(1,1,0,-1)(1,1,0,-1)(0,-1,-2,1)(0,-2,1,1)(-2,1,1,0)&(1,0,-1,0)(1,0,-1,0)(0,-1,1,0)(0,-1,-2,1)(-2,2,3,-1)\\\hline

\hline
CICY 7435: {\tiny $\left(\begin{array}{l}\mathbb{P}^1\\\mathbb{P}^1\\\mathbb{P}^1\\\mathbb{P}^7\end{array}\right|\left.\begin{array}{lllllll}1&1&0&0&0&0&0\\0&0&1&1&0&0&0\\0&0&0&0&1&1&0\\1&1&1&1&1&1&2\\\end{array}\right)^{4,44}_{-80}$}&$\mathbb{Z}_2\times\mathbb{Z}_2$\\\hline\hline
(2,1,1,-1)(2,1,-3,0)(-1,0,1,0)(-1,0,1,0)(-2,-2,0,1)&(2,1,1,-1)(2,-3,1,0)(-1,1,0,0)(-1,1,0,0)(-2,0,-2,1)\\\hline
(1,2,1,-1)(1,-1,0,0)(1,-1,0,0)(0,-2,-2,1)(-3,2,1,0)&(1,1,2,-1)(1,0,-1,0)(1,0,-1,0)(0,-2,-2,1)(-3,1,2,0)\\\hline
(1,2,1,-1)(1,2,-3,0)(0,-1,1,0)(0,-1,1,0)(-2,-2,0,1)&(1,1,2,-1)(1,-3,2,0)(0,1,-1,0)(0,1,-1,0)(-2,0,-2,1)\\\hline
\end{tabular}
\normalsize
\end{center}\vskip -0.5cm
\end{table}
\begin{table}[!h]
\begin{center}
\scriptsize
\begin{tabular}{|l|l|}
\hline
CICY 7862: {\tiny $\left(\begin{array}{l}\mathbb{P}^1\\\mathbb{P}^1\\\mathbb{P}^1\\\mathbb{P}^1\end{array}\right|\left.\begin{array}{l}2\\2\\2\\2\\\end{array}\right)^{4,68}_{-128}$}&$\mathbb{Z}_2\times\mathbb{Z}_2$\\\hline\hline
(1,-3,0,2)(0,1,0,-1)(0,1,0,-1)(0,0,-1,1)(-1,1,1,-1)&(1,-1,-1,1)(1,-2,0,1)(0,0,-1,1)(-1,2,2,-3)(-1,1,0,0)\\\hline
(1,-1,-1,1)(1,-1,-1,1)(0,1,2,-3)(-1,1,-1,1)(-1,0,1,0)&(1,0,-2,-1)(1,-2,1,2)(0,0,1,-1)(-1,1,0,0)(-1,1,0,0)\\\hline
(1,0,-2,-1)(1,-2,2,1)(0,0,1,-1)(-1,1,0,0)(-1,1,-1,1)&(1,0,-2,1)(1,-2,0,1)(0,1,1,-2)(-1,1,1,-1)(-1,0,0,1)\\\hline
(1,0,-2,1)(1,-2,1,0)(0,1,1,-2)(-1,1,0,0)(-1,0,0,1)&(1,0,-1,0)(1,-3,2,0)(0,1,0,-1)(0,1,0,-1)(-2,1,-1,2)\\\hline
(1,0,-3,0)(1,-2,3,0)(0,0,1,-1)(0,0,1,-1)(-2,2,-2,2)&(1,0,-1,0)(1,-1,2,-2)(1,-2,1,0)(0,1,-1,0)(-3,2,-1,2)\\\hline
\hline
CICY 5256: {\tiny $\left(\begin{array}{l}\mathbb{P}^1\\\mathbb{P}^1\\\mathbb{P}^1\\\mathbb{P}^1\\\mathbb{P}^3\end{array}\right|\left.\begin{array}{llll}1&1&0&0\\2&0&0&0\\0&0&1&1\\0&0&1&1\\1&1&1&1\\\end{array}\right)^{5,29}_{-48}$}&$\mathbb{Z}_2$\\\hline\hline
(1,-2,0,1,0)(0,1,1,1,-1)(0,1,-1,0,0)(0,0,1,-2,0)(-1,0,-1,0,1)&(1,1,0,1,-1)(1,-2,0,0,0)(0,1,1,-2,0)(-1,1,0,1,0)(-1,-1,-1,0,1)\\\hline
(1,1,0,1,-1)(1,0,1,-2,0)(0,-1,0,1,0)(0,-1,-1,0,1)(-2,1,0,0,0)&\\\hline
\hline
CICY 5256: {\tiny $\left(\begin{array}{l}\mathbb{P}^1\\\mathbb{P}^1\\\mathbb{P}^1\\\mathbb{P}^1\\\mathbb{P}^3\end{array}\right|\left.\begin{array}{llll}1&1&0&0\\2&0&0&0\\0&0&1&1\\0&0&1&1\\1&1&1&1\\\end{array}\right)^{5,29}_{-48}$}&$\mathbb{Z}_2\times\mathbb{Z}_2$\\\hline\hline
(1,1,0,1,-1)(0,1,-2,-2,1)(0,0,1,-1,0)(0,-2,1,1,0)(-1,0,0,1,0)&(1,0,-2,1,0)(1,-2,1,0,0)(0,1,1,-2,0)(-1,1,0,0,0)(-1,0,0,1,0)\\\hline
(1,1,-2,0,0)(1,-2,0,1,0)(0,1,1,-2,0)(-1,0,1,0,0)(-1,0,0,1,0)&(1,1,0,1,-1)(1,-2,0,1,0)(0,1,-2,-2,1)(-1,0,1,0,0)(-1,0,1,0,0)\\\hline
(1,1,0,1,-1)(1,-2,1,0,0)(0,1,-2,-2,1)(-1,0,1,0,0)(-1,0,0,1,0)&(1,0,-2,1,0)(1,-2,1,0,0)(0,1,0,-1,0)(0,0,1,-1,0)(-2,1,0,1,0)\\\hline
(1,0,-1,0,0)(1,-2,1,0,0)(0,1,1,-2,0)(0,0,-1,1,0)(-2,1,0,1,0)&(1,0,-1,0,0)(1,-2,0,1,0)(0,1,1,-2,0)(0,1,-1,0,0)(-2,0,1,1,0)\\\hline
(1,0,-2,1,0)(1,-1,0,0,0)(0,1,1,-2,0)(0,-1,1,0,0)(-2,1,0,1,0)&(1,0,0,-1,0)(1,0,-2,1,0)(0,1,0,-1,0)(0,-2,1,1,0)(-2,1,1,0,0)\\\hline
(1,0,-1,0,0)(1,0,-1,0,0)(0,1,1,1,-1)(0,1,1,-2,0)(-2,-2,0,1,1)&(1,0,0,-1,0)(1,0,-2,1,0)(0,1,1,1,-1)(0,1,0,-1,0)(-2,-2,1,0,1)\\\hline
(1,0,1,1,-1)(1,0,-2,1,0)(0,1,0,-1,0)(0,1,0,-1,0)(-2,-2,1,0,1)&(1,0,1,1,-1)(1,0,-1,0,0)(0,1,1,-2,0)(0,1,-1,0,0)(-2,-2,0,1,1)\\\hline
(1,1,-2,0,0)(1,-1,0,0,0)(0,1,1,1,-1)(0,1,0,-1,0)(-2,-2,1,0,1)&(1,1,-2,0,0)(1,0,1,-2,0)(0,-1,1,0,0)(0,-1,0,1,0)(-2,1,0,1,0)\\\hline
(1,1,-2,0,0)(1,0,1,1,-1)(1,0,0,-1,0)(-1,1,0,0,0)(-2,-2,1,0,1)&\\\hline
\hline
CICY 5452: {\tiny $\left(\begin{array}{l}\mathbb{P}^1\\\mathbb{P}^1\\\mathbb{P}^1\\\mathbb{P}^1\\\mathbb{P}^3\end{array}\right|\left.\begin{array}{llll}1&1&0&0\\0&0&1&1\\2&0&0&0\\0&0&2&0\\1&1&1&1\\\end{array}\right)^{5,29}_{-48}$}&$\mathbb{Z}_2$\\\hline\hline
(1,1,0,-2,0)(1,0,1,1,-1)(0,0,-1,1,0)(0,-1,-1,0,1)(-2,0,1,0,0)&(1,0,1,1,-1)(1,0,-2,0,0)(0,1,1,-2,0)(-1,0,0,1,0)(-1,-1,0,0,1)\\\hline
(1,1,-2,0,0)(0,1,1,1,-1)(0,0,1,-1,0)(0,-2,0,1,0)(-1,0,0,-1,1)&(1,0,-2,1,0)(0,1,1,1,-1)(0,1,0,-2,0)(0,-1,1,0,0)(-1,-1,0,0,1)\\\hline
\hline
CICY 5452: {\tiny $\left(\begin{array}{l}\mathbb{P}^1\\\mathbb{P}^1\\\mathbb{P}^1\\\mathbb{P}^1\\\mathbb{P}^3\end{array}\right|\left.\begin{array}{llll}1&1&0&0\\0&0&1&1\\2&0&0&0\\0&0&2&0\\1&1&1&1\\\end{array}\right)^{5,29}_{-48}$}&$\mathbb{Z}_2\times\mathbb{Z}_2$\\\hline\hline
(1,1,0,0,-1)(1,1,0,0,-1)(1,-2,0,0,1)(-1,0,-1,-1,1)(-2,0,1,1,0)&(1,1,0,0,-1)(1,1,0,0,-1)(1,-2,-1,1,1)(-1,0,1,-2,0)(-2,0,0,1,1)\\\hline
(1,1,0,1,-1)(1,0,1,-2,0)(1,-1,0,0,0)(-1,0,1,0,0)(-2,0,-2,1,1)&(1,1,0,0,-1)(1,1,0,0,-1)(0,0,-1,-2,1)(0,-2,1,1,0)(-2,0,0,1,1)\\\hline
(1,1,0,0,-1)(1,1,0,0,-1)(0,0,-2,-1,1)(0,-2,1,0,1)(-2,0,1,1,0)&(1,1,0,1,-1)(1,1,0,-2,0)(0,-1,1,0,0)(0,-1,1,0,0)(-2,0,-2,1,1)\\\hline
(1,1,0,0,-1)(1,1,0,0,-1)(0,-1,-1,-1,1)(0,-2,1,1,0)(-2,1,0,0,1)&(1,1,0,0,-1)(1,1,0,0,-1)(0,-1,-2,1,0)(0,-2,1,0,1)(-2,1,1,-1,1)\\\hline
(1,1,0,-2,0)(1,0,-1,0,0)(0,0,-1,1,0)(0,-2,1,1,0)(-2,1,1,0,0)&(1,1,-2,0,0)(1,0,0,-1,0)(0,0,1,-1,0)(0,-2,1,1,0)(-2,1,0,1,0)\\\hline
(1,1,0,-2,0)(1,0,-2,1,0)(0,-1,1,0,0)(0,-1,0,1,0)(-2,1,1,0,0)&(1,1,0,1,-1)(1,-1,0,0,0)(0,1,1,-2,0)(0,-1,1,0,0)(-2,0,-2,1,1)\\\hline
(1,1,0,-2,0)(1,-1,0,0,0)(0,1,1,1,-1)(0,-1,1,0,0)(-2,0,-2,1,1)&(1,1,0,-2,0)(1,-1,0,0,0)(0,1,-2,1,0)(0,-1,1,0,0)(-2,0,1,1,0)\\\hline
(1,1,-2,0,0)(1,-1,0,0,0)(0,1,1,-2,0)(0,-1,0,1,0)(-2,0,1,1,0)&(1,1,0,-2,0)(1,-2,1,0,0)(0,1,-1,0,0)(0,0,-1,1,0)(-2,0,1,1,0)\\\hline
(1,1,-2,0,0)(1,-2,0,1,0)(0,1,0,-1,0)(0,0,1,-1,0)(-2,0,1,1,0)&(1,0,1,-2,0)(1,0,-1,0,0)(0,1,1,1,-1)(0,-1,1,0,0)(-2,0,-2,1,1)\\\hline
(1,0,0,-1,0)(1,0,-2,1,0)(0,1,0,-1,0)(0,-2,1,1,0)(-2,1,1,0,0)&(1,0,-2,1,0)(1,-1,0,0,0)(0,1,1,-2,0)(0,-1,1,0,0)(-2,1,0,1,0)\\\hline
(1,0,-1,0,0)(1,-2,0,1,0)(0,1,1,-2,0)(0,1,-1,0,0)(-2,0,1,1,0)&(1,0,1,-2,0)(1,-2,0,1,0)(0,1,-1,0,0)(0,0,-1,1,0)(-2,1,1,0,0)\\\hline
(1,0,0,-1,0)(1,-2,0,1,0)(0,1,-2,1,0)(0,0,1,-1,0)(-2,1,1,0,0)&(1,-1,0,0,0)(1,-1,0,0,0)(0,1,1,1,-1)(0,1,1,-2,0)(-2,0,-2,1,1)\\\hline
(1,1,1,0,-1)(1,1,-2,0,0)(0,-2,1,-2,1)(-1,0,0,1,0)(-1,0,0,1,0)&(1,1,1,0,-1)(1,0,-2,1,0)(0,-2,1,-2,1)(-1,1,0,0,0)(-1,0,0,1,0)\\\hline
(1,1,0,-2,0)(1,0,-2,1,0)(0,-2,1,1,0)(-1,1,0,0,0)(-1,0,1,0,0)&(1,1,-2,0,0)(1,0,1,1,-1)(0,-2,1,-2,1)(-1,1,0,0,0)(-1,0,0,1,0)\\\hline
(1,1,-2,0,0)(1,0,1,-2,0)(0,-2,1,1,0)(-1,1,0,0,0)(-1,0,0,1,0)&(1,1,-2,0,0)(1,-2,0,1,0)(0,1,1,-2,0)(-1,0,1,0,0)(-1,0,0,1,0)\\\hline
(1,0,1,1,-1)(1,0,-2,1,0)(0,-2,1,-2,1)(-1,1,0,0,0)(-1,1,0,0,0)&(1,0,-2,1,0)(1,-2,1,0,0)(0,1,1,-2,0)(-1,1,0,0,0)(-1,0,0,1,0)\\\hline
(1,1,1,0,-1)(0,1,-2,1,0)(0,-1,0,1,0)(0,-2,1,-2,1)(-1,1,0,0,0)&(1,0,1,1,-1)(0,1,0,-1,0)(0,1,-2,1,0)(0,-2,1,-2,1)(-1,0,0,1,0)\\\hline\end{tabular}
\normalsize
\end{center}\vskip -0.5cm
\end{table}
\begin{table}[!h]
\begin{center}
\scriptsize
\begin{tabular}{|l|l|}
\hline
CICY 6947: {\tiny $\left(\begin{array}{l}\mathbb{P}^1\\\mathbb{P}^1\\\mathbb{P}^1\\\mathbb{P}^1\\\mathbb{P}^7\end{array}\right|\left.\begin{array}{llllllll}1&1&0&0&0&0&0&0\\0&0&1&1&0&0&0&0\\0&0&0&0&1&1&0&0\\0&0&0&0&0&0&1&1\\1&1&1&1&1&1&1&1\\\end{array}\right)^{5,37}_{-64}$}&$\mathbb{Z}_2$\\\hline\hline
(1,0,1,1,-1)(1,0,1,1,-1)(1,-2,0,-2,1)(-1,1,-2,-1,1)(-2,1,0,1,0)&(1,1,0,1,-1)(1,1,0,1,-1)(0,1,-2,-2,1)(0,-2,1,1,0)(-2,-1,1,-1,1)\\\hline
(1,0,1,1,-1)(1,0,1,1,-1)(0,0,-1,-2,1)(0,-1,-1,-1,1)(-2,1,0,1,0)&(1,1,0,1,-1)(1,1,0,1,-1)(0,-2,1,1,0)(-1,0,0,-2,1)(-1,0,-1,-1,1)\\\hline
\hline
CICY 6947: {\tiny $\left(\begin{array}{l}\mathbb{P}^1\\\mathbb{P}^1\\\mathbb{P}^1\\\mathbb{P}^1\\\mathbb{P}^7\end{array}\right|\left.\begin{array}{llllllll}1&1&0&0&0&0&0&0\\0&0&1&1&0&0&0&0\\0&0&0&0&1&1&0&0\\0&0&0&0&0&0&1&1\\1&1&1&1&1&1&1&1\\\end{array}\right)^{5,37}_{-64}$}&$\mathbb{Z}_2\times\mathbb{Z}_2$\\\hline\hline
(1,1,-2,-2,1)(1,0,1,1,-1)(1,0,1,1,-1)(-1,-2,0,-1,1)(-2,1,0,1,0)&(1,1,0,1,-1)(1,1,0,1,-1)(0,-1,-2,-1,1)(0,-2,1,1,0)(-2,1,1,-2,1)\\\hline
(1,1,0,-2,0)(1,0,-1,0,0)(0,0,-1,1,0)(0,-2,1,1,0)(-2,1,1,0,0)&(1,1,-2,0,0)(1,0,0,-1,0)(0,0,1,-1,0)(0,-2,1,1,0)(-2,1,0,1,0)\\\hline
(1,1,0,-2,0)(1,0,-2,1,0)(0,-1,1,0,0)(0,-1,0,1,0)(-2,1,1,0,0)&(1,1,-2,0,0)(1,0,1,-2,0)(0,-1,1,0,0)(0,-1,0,1,0)(-2,1,0,1,0)\\\hline
(1,1,0,-2,0)(1,-1,0,0,0)(0,1,-2,1,0)(0,-1,1,0,0)(-2,0,1,1,0)&(1,1,-2,0,0)(1,-1,0,0,0)(0,1,1,-2,0)(0,-1,0,1,0)(-2,0,1,1,0)\\\hline
(1,1,0,-2,0)(1,-2,1,0,0)(0,1,-1,0,0)(0,0,-1,1,0)(-2,0,1,1,0)&(1,1,-2,0,0)(1,-2,0,1,0)(0,1,0,-1,0)(0,0,1,-1,0)(-2,0,1,1,0)\\\hline
(1,0,1,-2,0)(1,0,-1,0,0)(0,1,-1,0,0)(0,-2,1,1,0)(-2,1,0,1,0)&(1,0,0,-1,0)(1,0,-2,1,0)(0,1,0,-1,0)(0,-2,1,1,0)(-2,1,1,0,0)\\\hline
(1,0,1,-2,0)(1,-1,0,0,0)(0,1,-2,1,0)(0,-1,0,1,0)(-2,1,1,0,0)&(1,0,-2,1,0)(1,-1,0,0,0)(0,1,1,-2,0)(0,-1,1,0,0)(-2,1,0,1,0)\\\hline
(1,0,0,-1,0)(1,-2,1,0,0)(0,1,0,-1,0)(0,1,-2,1,0)(-2,0,1,1,0)&(1,0,-1,0,0)(1,-2,0,1,0)(0,1,1,-2,0)(0,1,-1,0,0)(-2,0,1,1,0)\\\hline
(1,0,1,-2,0)(1,-2,0,1,0)(0,1,-1,0,0)(0,0,-1,1,0)(-2,1,1,0,0)&(1,0,0,-1,0)(1,-2,0,1,0)(0,1,-2,1,0)(0,0,1,-1,0)(-2,1,1,0,0)\\\hline
(1,0,-1,0,0)(1,-2,1,0,0)(0,1,1,-2,0)(0,0,-1,1,0)(-2,1,0,1,0)&(1,0,-2,1,0)(1,-2,1,0,0)(0,1,0,-1,0)(0,0,1,-1,0)(-2,1,0,1,0)\\\hline
(1,1,0,-2,0)(1,0,-2,1,0)(0,-2,1,1,0)(-1,1,0,0,0)(-1,0,1,0,0)&(1,1,-2,0,0)(1,0,1,-2,0)(0,-2,1,1,0)(-1,1,0,0,0)(-1,0,0,1,0)\\\hline
(1,1,0,-2,0)(1,-2,1,0,0)(0,1,-2,1,0)(-1,0,1,0,0)(-1,0,0,1,0)&(1,1,-2,0,0)(1,-2,0,1,0)(0,1,1,-2,0)(-1,0,1,0,0)(-1,0,0,1,0)\\\hline
(1,0,1,-2,0)(1,-2,0,1,0)(0,1,-2,1,0)(-1,1,0,0,0)(-1,0,1,0,0)&(1,0,-2,1,0)(1,-2,1,0,0)(0,1,1,-2,0)(-1,1,0,0,0)(-1,0,0,1,0)\\\hline
\hline
CICY 6732: {\tiny $\left(\begin{array}{l}\mathbb{P}^1\\\mathbb{P}^1\\\mathbb{P}^1\\\mathbb{P}^1\\\mathbb{P}^5\end{array}\right|\left.\begin{array}{llllll}1&1&0&0&0&0\\0&0&1&1&0&0\\0&0&0&0&1&1\\0&0&0&0&2&0\\1&1&1&1&1&1\\\end{array}\right)^{5,37}_{-64}$}&$\mathbb{Z}_2$\\\hline\hline
(1,1,1,0,-1)(1,0,-2,1,0)(1,-1,0,0,0)(-1,1,1,-1,0)(-2,-1,0,0,1)&(1,1,1,0,-1)(1,-1,1,-1,0)(1,-1,0,0,0)(-1,0,-2,0,1)(-2,1,0,1,0)\\\hline
(1,1,1,0,-1)(1,1,-1,1,-1)(0,0,-2,1,0)(0,-2,1,-1,1)(-2,0,1,-1,1)&(1,1,0,-2,0)(1,0,1,1,-1)(0,1,-1,0,0)(0,-1,0,1,0)(-2,-1,0,0,1)\\\hline
(1,1,-2,0,0)(1,0,1,1,-1)(0,1,0,-1,0)(0,-1,1,0,0)(-2,-1,0,0,1)&(1,1,1,0,-1)(1,-1,0,0,0)(0,1,0,-2,0)(0,0,-1,1,0)(-2,-1,0,1,1)\\\hline
(1,1,1,0,-1)(1,-1,0,0,0)(0,1,-2,1,0)(0,0,1,-1,0)(-2,-1,0,0,1)&(1,1,0,1,-1)(1,-1,1,-1,0)(0,1,-2,1,0)(0,0,1,-1,0)(-2,-1,0,0,1)\\\hline
(1,1,0,1,-1)(1,-1,0,0,0)(0,1,-2,0,0)(0,0,1,-1,0)(-2,-1,1,0,1)&(1,1,0,1,-1)(1,-1,1,-1,0)(0,0,-1,1,0)(0,-1,0,-2,1)(-2,1,0,1,0)\\\hline
(1,1,0,-2,0)(1,0,-1,0,0)(0,1,1,1,-1)(-1,0,0,1,0)(-1,-2,0,0,1)&(1,1,-2,0,0)(1,0,0,-1,0)(0,1,1,1,-1)(-1,0,1,0,0)(-1,-2,0,0,1)\\\hline
(1,1,1,0,-1)(1,0,0,-2,0)(0,0,-1,1,0)(-1,1,0,0,0)(-1,-2,0,1,1)&(1,1,1,0,-1)(1,0,-2,1,0)(0,0,1,-1,0)(-1,1,0,0,0)(-1,-2,0,0,1)\\\hline
(1,1,0,1,-1)(1,0,-2,1,0)(0,0,1,-1,0)(-1,1,1,-1,0)(-1,-2,0,0,1)&(1,1,0,1,-1)(1,0,-2,0,0)(0,0,1,-1,0)(-1,1,0,0,0)(-1,-2,1,0,1)\\\hline
(1,1,1,0,-1)(1,-1,1,-1,0)(0,1,-2,1,0)(-1,1,0,0,0)(-1,-2,0,0,1)&(1,1,1,0,-1)(1,-1,0,0,0)(0,0,1,-2,0)(-1,0,0,1,0)(-1,0,-2,1,1)\\\hline
(1,1,0,1,-1)(1,-2,0,1,0)(0,0,-1,1,0)(-1,1,1,-1,0)(-1,0,0,-2,1)&(1,1,1,0,-1)(1,-2,0,1,0)(0,-1,-2,0,1)(-1,1,1,-1,0)(-1,1,0,0,0)\\\hline
(1,1,1,0,-1)(0,0,1,-2,0)(0,-1,0,1,0)(0,-1,-2,1,1)(-1,1,0,0,0)&\\\hline
\hline
CICY 6770: {\tiny $\left(\begin{array}{l}\mathbb{P}^1\\\mathbb{P}^1\\\mathbb{P}^1\\\mathbb{P}^1\\\mathbb{P}^1\end{array}\right|\left.\begin{array}{ll}1&1\\1&1\\1&1\\2&0\\0&2\\\end{array}\right)^{5,37}_{-64}$}&$\mathbb{Z}_2$\\\hline\hline
(1,-2,0,0,0)(0,1,1,-2,0)(0,1,0,0,-1)(0,-1,-1,1,1)(-1,1,0,1,0)&(1,-2,0,0,0)(0,1,1,0,-2)(0,1,0,-1,0)(0,-1,-1,1,1)(-1,1,0,0,1)\\\hline
(1,-2,0,0,1)(0,1,0,0,-2)(0,1,-1,1,0)(0,0,0,-1,1)(-1,0,1,0,0)&(1,-2,0,1,0)(0,1,0,-2,0)(0,1,-1,0,1)(0,0,0,1,-1)(-1,0,1,0,0)\\\hline
(1,-2,0,0,0)(0,1,1,-2,0)(0,1,0,0,-1)(0,0,-1,1,1)(-1,0,0,1,0)&(1,-2,0,0,0)(0,1,1,0,-2)(0,1,0,-1,0)(0,0,-1,1,1)(-1,0,0,0,1)\\\hline
(1,-1,-1,1,1)(0,1,0,-2,1)(0,0,1,0,-2)(0,-1,0,1,0)(-1,1,0,0,0)&(1,-1,-1,1,1)(0,1,0,1,-2)(0,0,1,-2,0)(0,-1,0,0,1)(-1,1,0,0,0)\\\hline
(1,-1,0,0,0)(0,1,0,-2,1)(0,0,1,0,-2)(0,0,-1,1,1)(-1,0,0,1,0)&(1,-1,0,0,0)(0,1,0,1,-2)(0,0,1,-2,0)(0,0,-1,1,1)(-1,0,0,0,1)\\\hline
(1,-1,1,-1,1)(0,1,0,0,-2)(0,0,-1,1,0)(0,0,-1,0,1)(-1,0,1,0,0)&(1,-1,1,1,-1)(0,1,0,-2,0)(0,0,-1,1,0)(0,0,-1,0,1)(-1,0,1,0,0)\\\hline
(1,0,0,-2,0)(0,1,-2,0,1)(0,-1,1,1,-1)(0,-1,1,0,0)(-1,1,0,1,0)&(1,0,0,0,-2)(0,1,-2,1,0)(0,-1,1,0,0)(0,-1,1,-1,1)(-1,1,0,0,1)\\\hline
(1,0,0,-2,0)(0,1,-1,0,0)(0,0,1,0,-1)(0,-2,1,1,0)(-1,1,-1,1,1)&(1,0,0,0,-2)(0,1,-1,0,0)(0,0,1,-1,0)(0,-2,1,0,1)(-1,1,-1,1,1)\\\hline
\end{tabular}
\normalsize
\end{center}\vskip -0.5cm
\end{table}
\begin{table}[!h]
\begin{center}
\scriptsize
\begin{tabular}{|l|l|}
\hline
CICY 6777: {\tiny $\left(\begin{array}{l}\mathbb{P}^1\\\mathbb{P}^1\\\mathbb{P}^1\\\mathbb{P}^1\\\mathbb{P}^3\end{array}\right|\left.\begin{array}{llll}1&1&0&0\\0&0&0&2\\0&0&2&0\\2&0&0&0\\1&1&1&1\\\end{array}\right)^{5,37}_{-64}$}&$\mathbb{Z}_2$\\\hline\hline
(1,1,1,0,-1)(1,0,-2,-1,1)(1,-2,0,-1,1)(-1,1,1,1,-1)(-2,0,0,1,0)&(1,1,0,1,-1)(1,0,-1,0,0)(0,0,1,-1,0)(0,-2,-1,0,1)(-2,1,1,0,0)\\\hline
(1,0,1,1,-1)(1,-1,0,0,0)(0,1,0,-1,0)(0,-1,-2,0,1)(-2,1,1,0,0)&(1,0,0,-1,0)(1,-1,-2,0,1)(0,1,1,1,-1)(0,-1,1,0,0)(-2,1,0,0,0)\\\hline
(1,0,0,-1,0)(1,-2,-1,0,1)(0,1,1,1,-1)(0,1,-1,0,0)(-2,0,1,0,0)&(1,1,1,0,-1)(0,1,-1,0,0)(0,0,1,-2,0)(0,-2,-1,1,1)(-1,0,0,1,0)\\\hline
(1,1,0,1,-1)(0,1,1,-2,0)(0,0,-1,1,0)(0,-2,-1,0,1)(-1,0,1,0,0)&(1,1,-1,-1,0)(0,1,1,1,-1)(0,0,-1,-2,1)(0,-2,1,1,0)(-1,0,0,1,0)\\\hline
(1,1,1,0,-1)(0,1,0,-2,0)(0,-1,1,0,0)(0,-1,-2,1,1)(-1,0,0,1,0)&(1,0,1,1,-1)(0,1,1,-2,0)(0,-1,0,1,0)(0,-1,-2,0,1)(-1,1,0,0,0)\\\hline
(1,-1,1,-1,0)(0,1,1,1,-1)(0,1,-2,1,0)(0,-1,0,-2,1)(-1,0,0,1,0)&\\\hline
\hline
CICY 6890: {\tiny $\left(\begin{array}{l}\mathbb{P}^1\\\mathbb{P}^1\\\mathbb{P}^1\\\mathbb{P}^1\\\mathbb{P}^4\end{array}\right|\left.\begin{array}{lllll}1&1&0&0&0\\0&0&1&1&0\\0&0&0&0&2\\0&0&2&0&0\\1&1&1&1&1\\\end{array}\right)^{5,37}_{-64}$}&$\mathbb{Z}_2$\\\hline\hline
(1,1,1,0,-1)(1,0,-1,0,0)(1,-2,0,1,0)(-1,1,1,-1,0)(-2,0,-1,0,1)&(1,1,1,0,-1)(1,0,-1,0,0)(0,0,1,-2,0)(0,-1,0,1,0)(-2,0,-1,1,1)\\\hline
(1,1,0,1,-1)(1,0,1,-2,0)(0,0,-1,1,0)(0,-1,1,0,0)(-2,0,-1,0,1)&(1,1,-1,-1,0)(1,0,1,1,-1)(0,0,-1,-2,1)(0,-1,0,1,0)(-2,0,1,1,0)\\\hline
(1,1,1,0,-1)(1,-1,1,1,-1)(0,1,-2,-1,1)(0,-2,0,1,0)(-2,1,0,-1,1)&(1,1,0,1,-1)(1,-2,1,0,0)(0,1,-1,0,0)(0,0,1,-1,0)(-2,0,-1,0,1)\\\hline
(1,0,1,1,-1)(1,0,-1,0,0)(0,1,0,-1,0)(0,-2,1,0,0)(-2,1,-1,0,1)&(1,1,1,0,-1)(1,0,0,-2,0)(0,-1,0,1,0)(-1,0,1,0,0)(-1,0,-2,1,1)\\\hline
(1,1,1,0,-1)(1,0,-2,1,0)(0,-2,-1,0,1)(-1,1,1,-1,0)(-1,0,1,0,0)&(1,1,1,0,-1)(1,-2,0,1,0)(0,1,0,-1,0)(-1,0,1,0,0)(-1,0,-2,0,1)\\\hline
(1,0,1,1,-1)(1,0,-2,1,0)(0,-1,0,1,0)(-1,1,1,-1,0)(-1,0,0,-2,1)&(1,0,1,-2,0)(1,-1,0,0,0)(0,1,1,1,-1)(-1,0,0,1,0)(-1,0,-2,0,1)\\\hline
(1,0,1,1,-1)(1,-2,0,1,0)(0,1,0,-1,0)(-1,1,1,-1,0)(-1,0,-2,0,1)&(1,0,1,1,-1)(1,-2,0,0,0)(0,1,0,-1,0)(-1,1,-2,0,1)(-1,0,1,0,0)\\\hline
(1,0,0,-1,0)(1,-2,1,0,0)(0,1,1,1,-1)(-1,1,0,0,0)(-1,0,-2,0,1)&(1,1,1,0,-1)(0,1,0,-2,0)(0,0,-1,1,0)(0,-2,-1,1,1)(-1,0,1,0,0)\\\hline
\hline
CICY 7447: {\tiny $\left(\begin{array}{l}\mathbb{P}^1\\\mathbb{P}^1\\\mathbb{P}^1\\\mathbb{P}^1\\\mathbb{P}^1\end{array}\right|\left.\begin{array}{ll}1&1\\1&1\\1&1\\1&1\\1&1\\\end{array}\right)^{5,45}_{-80}$}&$\mathbb{Z}_2\times\mathbb{Z}_2$\\\hline\hline
(0,1,0,-2,1)(0,1,-2,1,0)(0,0,1,1,-2)(0,-1,1,0,0)(0,-1,0,0,1)&(1,-2,0,0,1)(0,1,-2,0,1)(0,0,1,1,-2)(0,0,1,-1,0)(-1,1,0,0,0)\\\hline
(1,-2,0,0,1)(0,1,0,1,-2)(0,0,1,-2,1)(0,0,-1,0,1)(-1,1,0,1,-1)&(1,-2,-1,1,1)(0,1,1,-2,0)(0,1,-1,0,0)(0,0,1,1,-2)(-1,0,0,0,1)\\\hline
\hline
CICY 7487: {\tiny $\left(\begin{array}{l}\mathbb{P}^1\\\mathbb{P}^1\\\mathbb{P}^1\\\mathbb{P}^1\\\mathbb{P}^1\end{array}\right|\left.\begin{array}{ll}0&2\\1&1\\1&1\\1&1\\1&1\\\end{array}\right)^{5,45}_{-80}$}&$\mathbb{Z}_2\times\mathbb{Z}_2$\\\hline\hline
(1,-2,0,0,1)(0,1,-2,0,1)(0,0,1,1,-2)(0,0,1,-1,0)(-1,1,0,0,0)&(1,-2,0,0,1)(0,1,0,1,-2)(0,0,1,-2,1)(0,0,-1,0,1)(-1,1,0,1,-1)\\\hline
(1,0,-2,0,1)(1,-2,1,0,0)(0,1,1,0,-2)(-1,1,0,0,0)(-1,0,0,0,1)&(1,-1,-1,0,1)(1,-2,0,0,1)(0,1,0,1,-2)(0,1,0,-1,0)(-2,1,1,0,0)\\\hline
(1,-1,-1,1,0)(1,-1,-1,0,1)(0,1,0,-2,1)(0,0,1,1,-2)(-2,1,1,0,0)&(1,-1,0,0,0)(1,-1,-1,0,1)(0,1,0,-2,1)(0,0,1,1,-2)(-2,1,0,1,0)\\\hline
(1,-1,1,-1,0)(1,-1,-2,1,1)(0,1,1,-2,0)(0,0,0,1,-1)(-2,1,0,1,0)&(1,-1,1,-1,0)(1,-1,-2,1,1)(0,1,1,0,-2)(0,0,0,-1,1)(-2,1,0,1,0)\\\hline
(1,0,-2,0,1)(1,-2,0,1,0)(0,1,0,-1,0)(0,0,1,0,-1)(-2,1,1,0,0)&(1,0,-2,0,1)(1,-2,0,1,0)(0,1,0,0,-1)(0,0,1,-1,0)(-2,1,1,0,0)\\\hline
(1,0,-2,0,1)(1,-2,1,0,0)(0,1,0,-1,0)(0,0,1,0,-1)(-2,1,0,1,0)&(1,0,-2,0,1)(1,-2,1,0,0)(0,1,0,0,-1)(0,0,1,-1,0)(-2,1,0,1,0)\\\hline
(1,0,-2,0,1)(1,-2,1,0,0)(0,1,0,0,-1)(0,0,1,0,-1)(-2,1,0,0,1)&(1,0,-2,0,1)(1,-2,1,1,-1)(0,1,0,-1,0)(0,0,1,0,-1)(-2,1,0,0,1)\\\hline
(1,0,-2,0,1)(1,-2,1,1,-1)(0,1,0,0,-1)(0,0,1,-1,0)(-2,1,0,0,1)&(1,0,-1,0,0)(1,-2,0,0,1)(0,1,1,-2,0)(0,0,0,1,-1)(-2,1,0,1,0)\\\hline
(1,0,-1,0,0)(1,-2,0,0,1)(0,1,1,0,-2)(0,0,0,-1,1)(-2,1,0,1,0)&(1,0,-1,0,0)(1,-2,1,0,0)(0,1,0,-2,1)(0,0,0,1,-1)(-2,1,0,1,0)\\\hline
(1,0,-1,0,0)(1,-2,1,0,0)(0,1,1,-2,0)(0,0,-1,1,0)(-2,1,0,1,0)&(1,0,-1,0,0)(1,-2,1,0,0)(0,1,1,-2,0)(0,0,0,1,-1)(-2,1,-1,1,1)\\\hline
(1,0,-1,0,0)(1,-2,0,0,1)(0,1,0,1,-2)(0,1,0,-1,0)(-2,0,1,0,1)&(1,0,-1,0,0)(1,-2,0,0,1)(0,1,1,0,-2)(0,1,-1,0,0)(-2,0,1,0,1)\\\hline
(1,0,-1,0,0)(1,-2,1,-1,1)(0,1,0,1,-2)(0,1,0,-1,0)(-2,0,0,1,1)&(1,0,0,-1,0)(1,0,-2,0,1)(1,-2,0,1,0)(-1,1,1,0,-1)(-2,1,1,0,0)\\\hline
\hline
CICY 6828: {\tiny $\left(\begin{array}{l}\mathbb{P}^1\\\mathbb{P}^1\\\mathbb{P}^1\\\mathbb{P}^3\end{array}\right|\left.\begin{array}{lll}0&0&2\\1&1&0\\1&1&0\\1&1&2\\\end{array}\right)^{4,36}_{-64}$}&$\mathbb{Z}_2\times\mathbb{Z}_2$\\\hline\hline
(1,0,2,-1)(1,-3,0,0)(0,1,-1,0)(0,1,-1,0)(-2,1,0,1)&(1,2,0,-1)(1,-3,2,0)(0,1,-1,0)(0,1,-1,0)(-2,-1,0,1)\\\hline
(2,-2,3,-1)(0,1,-1,0)(0,1,-1,0)(-1,0,1,0)(-1,0,-2,1)&(2,1,2,-1)(0,1,-3,0)(0,-2,-1,1)(-1,0,1,0)(-1,0,1,0)\\\hline
\end{tabular}
\normalsize
\end{center}\vskip -0.5cm
\end{table}



\clearpage
\begin{thebibliography}{99}

\bibitem{Green:1987mn}
  M.~B.~Green, J.~H.~Schwarz and E.~Witten,
  ``Superstring theory. Vol. 2: Loop amplitudes, anomalies and phenomenology,''
{\it  Cambridge, Uk: Univ. Pr. ( 1987) 596 P. ( Cambridge Monographs On Mathematical Physics)}

\bibitem{Braun:2005ux}
  V.~Braun, Y.~H.~He, B.~A.~Ovrut and T.~Pantev,
  ``A Heterotic standard model,''
  Phys.\ Lett.\  B {\bf 618}, 252 (2005)
  [arXiv:hep-th/0501070].

\bibitem{Braun:2005bw}
  V.~Braun, Y.~H.~He, B.~A.~Ovrut and T.~Pantev,
  ``A Standard model from the E(8) x E(8) heterotic superstring,''
  JHEP {\bf 0506}, 039 (2005)
  [arXiv:hep-th/0502155].

\bibitem{Braun:2005zv}
  V.~Braun, Y.~H.~He, B.~A.~Ovrut and T.~Pantev,
  ``Vector bundle extensions, sheaf cohomology, and the heterotic standard
  model,''
  Adv.\ Theor.\ Math.\ Phys.\  {\bf 10}, 4 (2006)
  [arXiv:hep-th/0505041].

\bibitem{Bouchard:2005ag}
  V.~Bouchard and R.~Donagi,
  ``An SU(5) heterotic standard model,''
  Phys.\ Lett.\  B {\bf 633}, 783 (2006)
  [arXiv:hep-th/0512149].

\bibitem{Braun:2005nv}
  V.~Braun, Y.~H.~He, B.~A.~Ovrut and T.~Pantev,
  ``The Exact MSSM spectrum from string theory,''
  JHEP {\bf 0605}, 043 (2006)
  [arXiv:hep-th/0512177].
  
\bibitem{Bouchard:2006dn}
  V.~Bouchard, M.~Cvetic and R.~Donagi,
  ``Tri-linear couplings in an heterotic minimal supersymmetric standard model,''
  Nucl.\ Phys.\  B {\bf 745}, 62 (2006)
  [arXiv:hep-th/0602096].

\bibitem{Blumenhagen:2006ux}
  R.~Blumenhagen, S.~Moster and T.~Weigand,
  ``Heterotic GUT and standard model vacua from simply connected Calabi-Yau
  manifolds,''
  Nucl.\ Phys.\  B {\bf 751}, 186 (2006)
  [arXiv:hep-th/0603015].


\bibitem{Blumenhagen:2006wj}
  R.~Blumenhagen, S.~Moster, R.~Reinbacher and T.~Weigand,
  ``Massless Spectra of Three Generation U(N) Heterotic String Vacua,''
  JHEP {\bf 0705}, 041 (2007)
  [arXiv:hep-th/0612039].
  
\bibitem{Anderson:2007nc}
  L.~B.~Anderson, Y.~H.~He and A.~Lukas,
  ``Heterotic Compactification, An Algorithmic Approach,''
  JHEP {\bf 0707}, 049 (2007)
  [arXiv:hep-th/0702210].

\bibitem{Anderson:2008uw}
  L.~B.~Anderson, Y.~H.~He and A.~Lukas,
  ``Monad Bundles in Heterotic String Compactifications,''
  JHEP {\bf 0807}, 104 (2008)
  [arXiv:0805.2875 [hep-th]].

\bibitem{Anderson:2009mh}
  L.~B.~Anderson, J.~Gray, Y.~H.~He and A.~Lukas,
  ``Exploring Positive Monad Bundles And A New Heterotic Standard Model,''
  JHEP {\bf 1002}, 054 (2010)
  [arXiv:0911.1569 [hep-th]].

\bibitem{Buchmuller:2005jr}
  W.~Buchmuller, K.~Hamaguchi, O.~Lebedev, M.~Ratz,
  ``Supersymmetric standard model from the heterotic string,''
  Phys.\ Rev.\ Lett.\  {\bf 96}, 121602 (2006).
  [hep-ph/0511035].

\bibitem{Buchmuller:2006ik}
  W.~Buchmuller, K.~Hamaguchi, O.~Lebedev, M.~Ratz,
  ``Supersymmetric Standard Model from the Heterotic String (II),''
  Nucl.\ Phys.\  {\bf B785}, 149-209 (2007).
  [hep-th/0606187].

\bibitem{Lebedev:2006kn}
  O.~Lebedev, H.~P.~Nilles, S.~Raby, S.~Ramos-Sanchez, M.~Ratz, P.~K.~S.~Vaudrevange, A.~Wingerter,
  ``A Mini-landscape of exact MSSM spectra in heterotic orbifolds,''
  Phys.\ Lett.\  {\bf B645}, 88-94 (2007).
  [hep-th/0611095].

\bibitem{Kim:2007mt}
  J.~E.~Kim, J.~-H.~Kim, B.~Kyae,
  ``Superstring standard model from Z(12-I) orbifold compactification with and without exotics, and effective R-parity,''
  JHEP {\bf 0706 } (2007)  034.
  [hep-ph/0702278 [HEP-PH]].

\bibitem{Lebedev:2008un}
  O.~Lebedev, H.~P.~Nilles, S.~Ramos-Sanchez, M.~Ratz, P.~K.~S.~Vaudrevange,
  ``Heterotic mini-landscape. (II). Completing the search for MSSM vacua in a Z(6) orbifold,''
  Phys.\ Lett.\  {\bf B668}, 331-335 (2008).
  [arXiv:0807.4384 [hep-th]].

\bibitem{Nibbelink:2009sp}
  S.~G.~Nibbelink, J.~Held, F.~Ruehle, M.~Trapletti, P.~K.~S.~Vaudrevange,
  ``Heterotic Z(6-II) MSSM Orbifolds in Blowup,''
  JHEP {\bf 0903}, 005 (2009).
  [arXiv:0901.3059 [hep-th]].

\bibitem{Blaszczyk:2009in}
  M.~Blaszczyk, S.~G.~Nibbelink, M.~Ratz, F.~Ruehle, M.~Trapletti, P.~K.~S.~Vaudrevange,
  ``A Z2xZ2 standard model,''
  Phys.\ Lett.\  {\bf B683}, 340-348 (2010).
  [arXiv:0911.4905 [hep-th]].

\bibitem{Blaszczyk:2010db}
  M.~Blaszczyk, S.~G.~Nibbelink, F.~Ruehle, M.~Trapletti, P.~K.~S.~Vaudrevange,
  ``Heterotic MSSM on a Resolved Orbifold,''
  JHEP {\bf 1009}, 065 (2010).
  [arXiv:1007.0203 [hep-th]].

\bibitem{Kappl:2010yu}
  R.~Kappl, B.~Petersen, S.~Raby, M.~Ratz, R.~Schieren, P.~K.~S.~Vaudrevange,
  ``String-derived MSSM vacua with residual R symmetries,''
  Nucl.\ Phys.\  {\bf B847}, 325-349 (2011).
  [arXiv:1012.4574 [hep-th]].


\bibitem{Assel:2009xa}
  B.~Assel, K.~Christodoulides, A.~E.~Faraggi, C.~Kounnas and J.~Rizos,
  ``Exophobic Quasi-Realistic Heterotic String Vacua,''
  Phys.\ Lett.\  B {\bf 683} (2010) 306
  [arXiv:0910.3697 [hep-th]].

\bibitem{Christodoulides:2011zs}
  K.~Christodoulides, A.~E.~Faraggi, J.~Rizos,
  ``Top Quark Mass in Exophobic Pati-Salam Heterotic String Model,''  
  [arXiv:1104.2264 [hep-ph]].

\bibitem{Cleaver:2011ir}
  G.~Cleaver, A.~E.~Faraggi, J.~Greenwald, D.~Moore, K.~Pechan, E.~Remkus, T.~Renner,
  ``Investigation of Quasi--Realistic Heterotic String Models with Reduced Higgs Spectrum,''
  [arXiv:1105.0447 [hep-ph]].

\bibitem{He:2003tj}
  Y.~H.~He, B.~A.~Ovrut and R.~Reinbacher,
  ``The Moduli of reducible vector bundles,''
  JHEP {\bf 0403}, 043 (2004)
  [arXiv:hep-th/0306121].

\bibitem{Blumenhagen:2005ga}
  R.~Blumenhagen, G.~Honecker and T.~Weigand,
  ``Loop-corrected compactifications of the heterotic string with line
  bundles,''
  JHEP {\bf 0506}, 020 (2005)
  [arXiv:hep-th/0504232].

\bibitem{He:2009wi}
  Y.~H.~He, S.~J.~Lee and A.~Lukas,
  ``Heterotic Models from Vector Bundles on Toric Calabi-Yau Manifolds,''
  JHEP {\bf 1005}, 071 (2010)
  [arXiv:0911.0865 [hep-th]].

 \bibitem{monad_scan}
L.~B.~ Anderson, J.~ Gray, Y.~H.~He, S.~J.~Lee and A. Lukas, ``Compactifying on Complete Intersections'',
To appear.

 \bibitem{singular}
  G.~M.~Greuel, G.~Pfister, and H.~Schonemann, 
  ``Singular: a computer algebra system for
polynomial computations,Ó Centre for Computer Algebra, University of Kaiserslautern
(2001). Available at http://www.singular.uni-kl.de/ .
  
\bibitem{Gray:2008zs}
  J.~Gray, Y.~H.~He, A.~Ilderton and A.~Lukas,
  ``STRINGVACUA: A Mathematica Package for Studying Vacuum Configurations in
  String Phenomenology,''
  Comput.\ Phys.\ Commun.\  {\bf 180}, 107 (2009)
  [arXiv:0801.1508 [hep-th]].
  
 \bibitem{cicypackage} L.~B.~Anderson, J.~Gray, Y.~H.~He, S.~J.~Lee,
  and A.~Lukas, ``CICY package", based on methods described in
  arXiv:0911.1569, arXiv:0911.0865, arXiv:0805.2875, hep-th/0703249,
  hep-th/0702210".

\bibitem{Anderson:2010tc}
  L.~B.~Anderson, J.~Gray and B.~Ovrut,
  ``Yukawa Textures From Heterotic Stability Walls,''
  JHEP {\bf 1005}, 086 (2010)
  [arXiv:1001.2317 [hep-th]].

\bibitem{Kuriyama:2008pv}
  M.~Kuriyama, H.~Nakajima and T.~Watari,
  ``Theoretical Framework for R-parity Violation,''
  Phys.\ Rev.\  D {\bf 79}, 075002 (2009)
  [arXiv:0802.2584 [hep-ph]].

\bibitem{bigpaper}   L.~B.~Anderson, J.~Gray, A.~ Lukas, and E.~Palti, To Appear.

\bibitem{the_cy_list}
  The CALABI-YAU Home Page,
http://www.th.physik.uni-bonn.de/th/Supplements/cy.html

\bibitem{Gagnon:1994ek}
  M.~Gagnon and Q.~Ho-Kim,
  ``An Exhaustive list of complete intersection Calabi-Yau manifolds,''
  Mod.\ Phys.\ Lett.\  A {\bf 9}, 2235 (1994).
 
\bibitem{Braun:2010vc}
  V.~Braun,
  ``On Free Quotients of Complete Intersection Calabi-Yau Manifolds,''
  JHEP {\bf 1104}, 005 (2011)
  [arXiv:1003.3235 [hep-th]].

\bibitem{Anderson:2008ex}
  L.~B.~Anderson,
  ``Heterotic and M-theory Compactifications for String Phenomenology,''
  arXiv:0808.3621 [hep-th].

\bibitem{Donagi:2003tb}
  R.~Donagi, B.~A.~Ovrut, T.~Pantev and R.~Reinbacher,
  ``SU(4) instantons on Calabi-Yau threefolds with Z(2) x Z(2) fundamental
  group,''
  JHEP {\bf 0401}, 022 (2004)
  [arXiv:hep-th/0307273].

\bibitem{Lukas:1999nh}
  A.~Lukas and K.~S.~Stelle,
  ``Heterotic anomaly cancellation in five-dimensions,''
  JHEP {\bf 0001}, 010 (2000)
  [arXiv:hep-th/9911156].


\bibitem{Anderson:2010ty}
  L.~B.~Anderson, J.~Gray and B.~Ovrut,
  ``Transitions in the Web of Heterotic Vacua,''
  Fortschr. Phys.  {\bf 59}, No. 5-6, 327 (2011)
  arXiv:1012.3179 [hep-th].

\bibitem{Anderson:2009nt}
  L.~B.~Anderson, J.~Gray, A.~Lukas and B.~Ovrut,
  ``Stability Walls in Heterotic Theories,''
  JHEP {\bf 0909}, 026 (2009)
  [arXiv:0905.1748 [hep-th]].
  
\bibitem{Anderson:2009sw}
  L.~B.~Anderson, J.~Gray, A.~Lukas and B.~Ovrut,
  ``The Edge Of Supersymmetry: Stability Walls in Heterotic Theory,''
  Phys.\ Lett.\  B {\bf 677}, 190 (2009)
  [arXiv:0903.5088 [hep-th]].

\bibitem{Gmeiner:2005vz}
  F.~Gmeiner, R.~Blumenhagen, G.~Honecker, D.~Lust and T.~Weigand,
  ``One in a billion: MSSM-like D-brane statistics,''
  JHEP {\bf 0601}, 004 (2006)
  [arXiv:hep-th/0510170].
  
\bibitem{Heckman:2009mn}
  J.~J.~Heckman, A.~Tavanfar and C.~Vafa,
  ``The Point of E(8) in F-theory GUTs,''
  JHEP {\bf 1008}, 040 (2010)
  [arXiv:0906.0581 [hep-th]].
  

\bibitem{Dudas:2009hu}
  E.~Dudas and E.~Palti,
  ``Froggatt-Nielsen models from E(8) in F-theory GUTs,''
  JHEP {\bf 1001}, 127 (2010)
  [arXiv:0912.0853 [hep-th]].

\bibitem{Marsano:2009wr}
  J.~Marsano, N.~Saulina, S.~Schafer-Nameki,
  ``Compact F-theory GUTs with U(1) (PQ),''
  JHEP {\bf 1004 } (2010)  095.
  [arXiv:0912.0272 [hep-th]].


\bibitem{Dudas:2010zb}
  E.~Dudas, E.~Palti,
  ``On hypercharge flux and exotics in F-theory GUTs,''
  JHEP {\bf 1009 } (2010)  013.
  [arXiv:1007.1297 [hep-ph]].

\bibitem{Marsano:2010sq}
  J.~Marsano,
  ``Hypercharge Flux, Exotics, and Anomaly Cancellation in F-theory GUTs,''
  Phys.\ Rev.\ Lett.\  {\bf 106}, 081601 (2011)
  [arXiv:1011.2212 [hep-th]].


  
\bibitem{Anderson:2009ge}
  L.~B.~Anderson, J.~Gray, D.~Grayson, Y.~H.~He and A.~Lukas,
  ``Yukawa Couplings in Heterotic Compactification,''
  Commun.\ Math.\ Phys.\  {\bf 297} (2010) 95
  [arXiv:0904.2186 [hep-th]].

\bibitem{Dolan:2011iu}
  M.~J.~Dolan, J.~Marsano, N.~Saulina, S.~Schafer-Nameki,
  ``F-theory GUTs with U(1) Symmetries: Generalities and Survey,''  
  [arXiv:1102.0290 [hep-th]].

\bibitem{Kim:1994eu}
  J.~E.~Kim and H.~P.~Nilles,
  ``Symmetry principles toward solutions of the mu problem,''
  Mod.\ Phys.\ Lett.\  A {\bf 9}, 3575 (1994)
  [arXiv:hep-ph/9406296].


\bibitem{Froggatt:1978nt}
  C.~D.~Froggatt and H.~B.~Nielsen,
  ``Hierarchy of Quark Masses, Cabibbo Angles and CP Violation,''
  Nucl.\ Phys.\  B {\bf 147}, 277 (1979).
  

 

  

\bibitem{Kreuzer:2000qv}
  M.~Kreuzer and H.~Skarke,
  ``Reflexive polyhedra, weights and toric Calabi-Yau fibrations,''
  Rev.\ Math.\ Phys.\  {\bf 14}, 343 (2002)
  [arXiv:math/0001106].

\bibitem{Kreuzer:2000xy}
  M.~Kreuzer and H.~Skarke,
  ``Complete classification of reflexive polyhedra in four-dimensions,''
  Adv.\ Theor.\ Math.\ Phys.\  {\bf 4}, 1209 (2002)
  [arXiv:hep-th/0002240].

\bibitem{Kreuzer:2002uu}
  M.~Kreuzer and H.~Skarke,
  ``PALP: A Package for analyzing lattice polytopes with applications to toric
  geometry,''
  Comput.\ Phys.\ Commun.\  {\bf 157}, 87 (2004)
  [arXiv:math/0204356].

\bibitem{Blumenhagen:2010pv}
  R.~Blumenhagen, B.~Jurke, T.~Rahn and H.~Roschy,
  ``Cohomology of Line Bundles: A Computational Algorithm,''
  J.\ Math.\ Phys.\  {\bf 51}, 103525 (2010)
  [arXiv:1003.5217 [hep-th]].
  

  


  

  
 


  
\end{thebibliography}
\end{document}